\documentclass[fleqn,usenatbib,useAMS]{mnras}
\usepackage{amssymb,amsmath}
\usepackage{graphicx}
\usepackage{xcolor,natbib}
\usepackage{multirow}
\usepackage[T1]{fontenc}
\usepackage{ae,aecompl}
\usepackage{newtxtext, newtxmath}
\usepackage{float}
\usepackage{subfigure}
\usepackage{longtable}
\usepackage{enumitem}
\usepackage{longtable,listings}
\usepackage[flushleft]{threeparttable}
\usepackage{parcolumns}
\def\dif{\mathop{}\!\mathrm{d}}
\def\obj{SDSS J1609+1756}

\title[A sub-pc BBH system]{A sub-pc BBH system in SDSS J1609+1756 through optical QPOs in ZTF light curves}
\author[Zhang X. G.]{XueGuang Zhang$^{1}$
\thanks{Contact e-mail: \href{mailto:xgzhang@gxu.edu.cn}{xgzhang@gxu.edu.cn}}\\
$^{1}$Guangxi Key Laboratory for Relativistic Astrophysics, School of Physical Science and Technology, Guangxi University,
          Nanning, 530004, P. R. China}

\begin{document}
\label{firstpage}
\pagerange{\pageref{firstpage}--\pageref{lastpage}}

\maketitle

\begin{abstract} %%%about 245 words
	Optical quasi-periodic oscillations (QPOs) are the most preferred signs of sub-pc binary black hole (BBH) systems 
in AGN. In this manuscript, robust optical QPOs are reported in quasar SDSS J1609+1756 at $z=0.347$. In order to detect 
reliable optical QPOs, four different methods are applied to analyze the 4.45 years-long ZTF g/r/i-band light curves of \obj, 
direct fitting results by sine function, Generalized Lomb-Scargle periodogram, Auto-Cross Correlation Function and Weighted 
Wavelet Z-transform method. The Four different methods can lead to well determined reliable optical QPOs with periodicities 
$\sim340$ days with confidence levels higher than 5$\sigma$, to guarantee the robustness of the optical QPOs in \obj. 
Meanwhile, based on simulated light curves through CAR process to trace intrinsic AGN activities, confidence level higher 
than $3\sigma$ can be confirmed that the optical QPOs are not mis-detected in intrinsic AGN activities, re-confirming the 
robust optical QPOs and strongly indicating a central sub-pc BBH system in \obj. Furthermore, based on apparent red-shifted 
shoulders in broad Balmer emission lines in \obj, space separation of the expected central BBH system can be estimated to 
be smaller than $107\pm60$ light-days, accepted upper limit of total BH mass $\sim(1.03\pm0.22)\times10^8{\rm M_\odot}$. 
Therefore, to detect and report BBH system expected optical QPOs with periodicities around 1 year is efficiently practicable 
through ZTF light curves, and combining with peculiar broad line emission features, further clues should be given on space 
separations of BBH systems in broad line AGN in the near future.  
\end{abstract}


\section{Introduction}

%%first
	Binary black hole (BBH) systems on scale of sub-parsecs in central regions of active galactic nuclei (AGN), as well as 
dual core systems on scale of kpcs (or AGN pairs), are common as discussed in \citet{bb80, mk10, fg19, mj22, ws23}, considering 
galaxy merging as an essential process of galaxy formation and evolution \citep{kw93, sr98, lk04, md06, bf09, se14, rs17, bh19, 
mj21, yp22}. Meanwhile, in the manuscript, through discussions in more recent reviews in \citet{dv19, ch22}, a kpc dual core 
system means central two BHs are getting closer due to dynamical frictions, but a sub-pc BBH system means central two BHs are 
getting closer mainly due to emission of gravitational waves. Besides indicators for BBH systems and/or dual core systems by 
spectroscopic features as discussed in \citet{zw04, kz08, bl09, ss09, sl10, eb12, cs13, le16, wg17, dv19, zh21d} and by spatial 
resolved image properties as discussed in \citet{km03, rt09, pv10, ne17, kw20}, long-standing optical Quasi-Periodic Oscillations 
(QPOs) with periodicities around hundreds to thousands of days have been commonly accepted as the most preferred indicators for 
central BBH systems in AGN.

	Long-standing optical QPOs have been reported in AGN related to central BBH systems in the literature. In the known 
quasar PG 1302-102, \citet{gd15a, lg18, kp19} have shown detailed discussions on reliable 1800 days optical QPOs. Meanwhile, 
strong evidence have been reported to support optical QPOs in other individual AGN, such as 540 days QPOs in PSO 
J334.2028+01.4075\ in \citet{lg15}, 1500 days QPOs in SDSS J0159\ in \citet{zb16}, 1150 days QPOs in Mrk915\ in \citet{ss20}, 
1.2 years QPOs in Mrk 231\ in \citet{ky20}, 1607 days QPOs in SDSS J0252\ in \citet{lw21}, 6.4 years optical QPOs in SDSS 
J0752\ in \citet{zh22a}, 3.8 years optical QPOs in SDSS J1321\ in \citet{zh22c}, etc. Moreover, besides the optical QPOs 
reported in individual AGN, a sample of 111 candidates with optical QPOs have been reported in \citet{gd15} based on strong 
Keplerian periodic signals over a baseline of nine years, and a sample of 50 candidates with optical QPOs have been reported 
in \citet{cb16}.

%The BBH systems can 
%produce expected background gravitational wave signals at nano-Hz frequencies probed by the Pulsar 
%Timing Arrays \citep{fb90, de16, re16, ar15, ve16}. However, besides the reported QPOs to support 
%BBH systems, false periodicities have been discussed in quasar time-domain surveys, such as the 
%results in \citet{vu16} and in \citet{se18}. Therefore, it is necessary and meaningful to detect and 
%report more candidates of BBH systems. 

	While detecting BBH systems through optical QPOs, two important points have serious effects on reliability of BBH 
system expected QPOs. First, comparing with periodicities of detected optical QPOs, time durations of light curves are not 
longer enough to support reliabilities of the QPOs. Second, central AGN activities should lead to false optical QPOs, as well 
discussed in \citet{vu16, se18, zh22a, zh22c}. Thus, the reported confidence levels for reported optical QPOs through 
mathematical methods should be carefully re-checked. Currently, there are many public Sky Survey projects conveniently applied 
to search for long-standing optical QPOs. However, as the shown results in the largest sample of optical QPOs in \citet{gd15} 
and the other reported optical QPOs in individual AGN, the reported optical QPOs have periodicities commonly around 3.5 years 
(1500 days). Therefore, the Catalina Sky Survey (CSS, \citealt{dr09}) with longer time durations and moderate data quality of 
light curves is the preferred Sky Survey project for conveniently and systematically searching for optical QPOs with a few 
years long periodicities as the brilliant works in \citet{gd15}. Meanwhile, comparing with the CSS project, the other Sky 
Survey projects have some disadvantages for searching for optical QPOs with ears-long periodicities, the Zwicky Transient 
Facility (ZTF, \citealt{bk19, ds20}) sky survey has light curves with short time durations (only around 4.5 years), the Panoramic 
Survey Telescope And Rapid Response System (PanSTARRS, \citealt{fm20, mc20}) and the Sloan Digital Sky Survey Stripe82 (SDSS 
Stripe82, \citealt{bv08, ti21}) sky survey has light curves with large time steps, the All-Sky Automated Survey for Supernovae 
(ASAS-SN, \citealt{sp14, ks17}) has light curves with limits for bright galaxies, etc. However, considering the high quality 
light curves in ZTF sky survey, optical QPOs with shorter periodicities (smaller than or around 1 year) should be preferred 
to be detected only through ZTF light curves, which is the main objective of the manuscript.

	In this manuscript, a new BBH candidate is reported in SDSS J160911.25+175616.22 (=\obj), a blue quasar at redshift 
0.347, due to detected optical QPOs with periodicities about 340 days through more recent ZTF g/r/i-band light curves. Moreover, 
due to apparent red-shifted shoulders in broad Balmer emission lines, properties of peak separations of broad emission lines 
can be applied to determine limits of space separation of central BBH system in \obj. The manuscript is organized as follows. 
Section 2 presents main results on the long-term optical variabilities of \obj, and four different methods to detect robust 
optical QPOs in \obj. Section 3 gives the main discussions including statistical results to support the optical QPOs not from 
central intrinsic AGN activities in \obj, also including discussions on spectroscopic properties of \obj, and discussions on 
basic structure information of space separation of central BBH system in \obj. Section 4 gives final summary and main conclusions. 
In the manuscript, the cosmological parameters well discussed in \citet{hl13} have been adopted as 
$H_{0}=70{\rm km\cdot s}^{-1}{\rm Mpc}^{-1}$, $\Omega_{\Lambda}=0.7$ and $\Omega_{\rm m}=0.3$.

\begin{figure*}
\centering\includegraphics[width = 18cm,height=12cm]{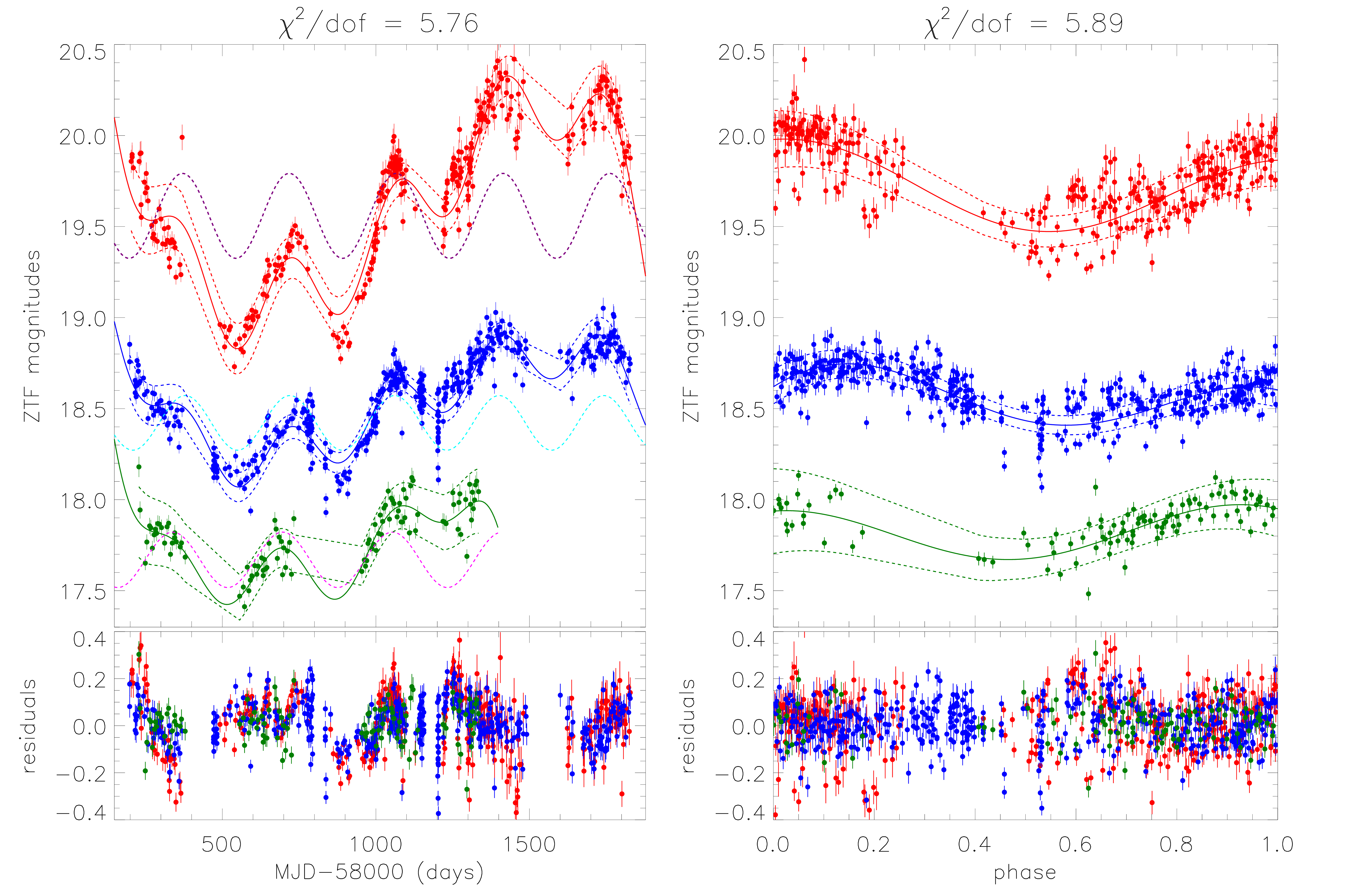}
\caption{Top left panel shows the ZTF g/r/i-band light curves and the best fitting results to the light curves by a sine function 
plus a five-degree polynomial function. Top right panel shows corresponding phase folded light curves and the best-fitting results 
by a sine function. Bottom panels show corresponding residuals calculated by light curves minus the best fitting results. In top 
right (left) panel, solid circles plus error bars in red, in blue and in dark green show the (folded) g-band light curve (plus 
0.5 magnitudes), the (folded) r-band light curve, and the (folded) i-band light curve (minus 0.5 magnitudes), respectively, solid 
and dashed lines in red, in blue and in dark green show the best fitting results and corresponding F-test technique determined 
$5\sigma$ confidence bands to the (folded) g-band light curve, to the (folded) r-band light curve and to the (folded) i-band light 
curve, respectively. In top left panel, dashed line in purple, in cyan and in magenta show the determined sine component included 
in the g-band light curve, in the r-band light curve and in the i-band light curve, respectively.}
\label{lmc}
\end{figure*}

\section{Optical QPOs in \obj}

\subsection{Long-term optical light curves of \obj}

%%2.1 - 1
	\obj~ is collected as target of the manuscript, due to two main reasons. First, as a candidate of off-nucleus AGN in 
\citet{wg21}, \obj~ has peculiar broad Balmer lines with shifted red shoulders which can be well explained by broad line 
emissions from central two independent BLRs related to a central BBH system, quite similar as what we have recently discussed 
in \citet{zh21d}. Second, after checking long-term variabilities of \obj, apparent optical QPOs can be detected to support an 
expected BBH system.

\begin{figure*}
\centering\includegraphics[width = 18cm,height=4cm]{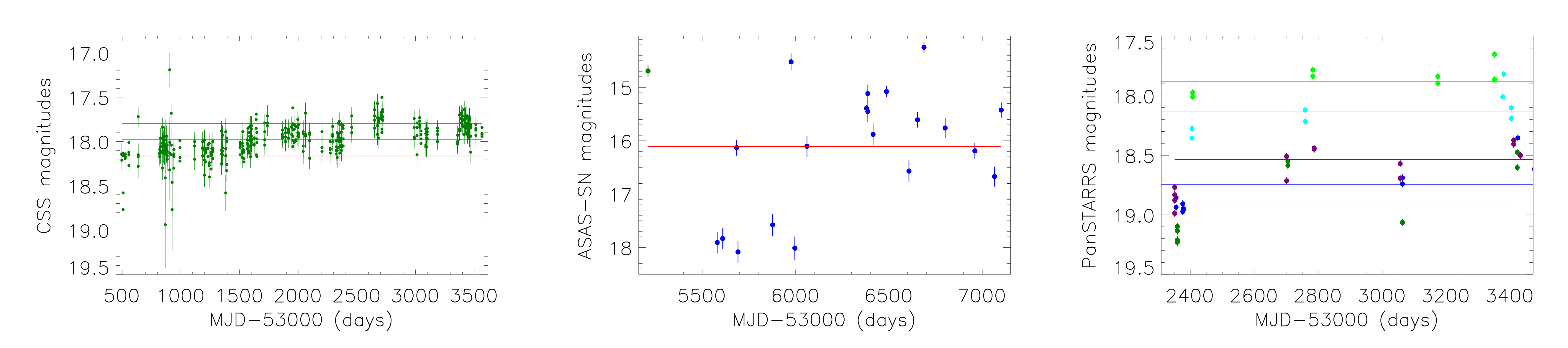}
\caption{Left panel shows the CSS light curve. Horizontal red lines show the mean value and corresponding 1RMS scatter bands 
of the light curve. Middle panel shows the ASAS-SN V-band (only one solid circle in dark green) and g-band (solid circles in 
blue) light curves. Horizontal red line shows the mean value of the g-band light curve. Right panel shows the PanSTARRS g-band 
(circles in in dark green), r-band (circles in blue), i-band (circles in purple), z-band (circles in green) and y-band (circles 
in cyan) light curves. Horizontal dark green line, blue line, purple line, green line and cyan line show the mean value of 
corresponding g-band, r-band, i-band, z-band, and y-band light curves. In middle and right panel, due to less number of data 
points, 1RMS scatter bands are not plotted to each light curve.}
\label{css}
\end{figure*}

%%2.1 - 2
	The 4.45 years-long ZTF g/r/i-band light curves of \obj~ are collected and shown in top left panel of Fig~\ref{lmc}, with 
MJD-58000 from 203 (Mar. 2018) to 1828 (Jul. 2022). Meanwhile, long-term light curves of \obj~ are also checked in CSS, PanSTARRS 
and ASAS-SN, and shown in Fig.~\ref{css} without apparent variabilities. There are 390 reliable data points included in the CSS 
light curve, however, as shown in left panel of Fig.~\ref{css}, more than 92\% of the 390 data points are lying between the range 
of mean magnitude plus/minus corresponding 1RMS scatters, indicating not apparent variabilities in the CSS light curve, probably 
due to lower quality of light curves. There are 1 reliable data point and 20 reliable data points included in the ASAS-SN V/g-band 
light curves shown in middle panel of Fig.~\ref{css}. However, due to he quite large time steps with mean value about 200 days, 
it is not appropriate to search QPOs in ASAS-SN light curves. Similar as ASAS-SN g-band light curve, there are only 10 data points, 
14 data points, 21 data points, 11 data points and 13 data points included in the PanSTARRS g/r/i/z/y-band light curves shown in 
right panel of Fig.~\ref{css}, respectively, with large mean time steps about 266 days, 217 days, 184 days, 20 days and 182 days. 
Therefore, the PanSTARRS light curves are not considered in the manuscript.

%%2.1 - 3
	Finally, the long-term ZTF light curves are mainly considered and there are no further discussions on variabilities from 
the other Sky Survey projects.

\begin{figure*}
\centering\includegraphics[width = 18cm,height=5cm]{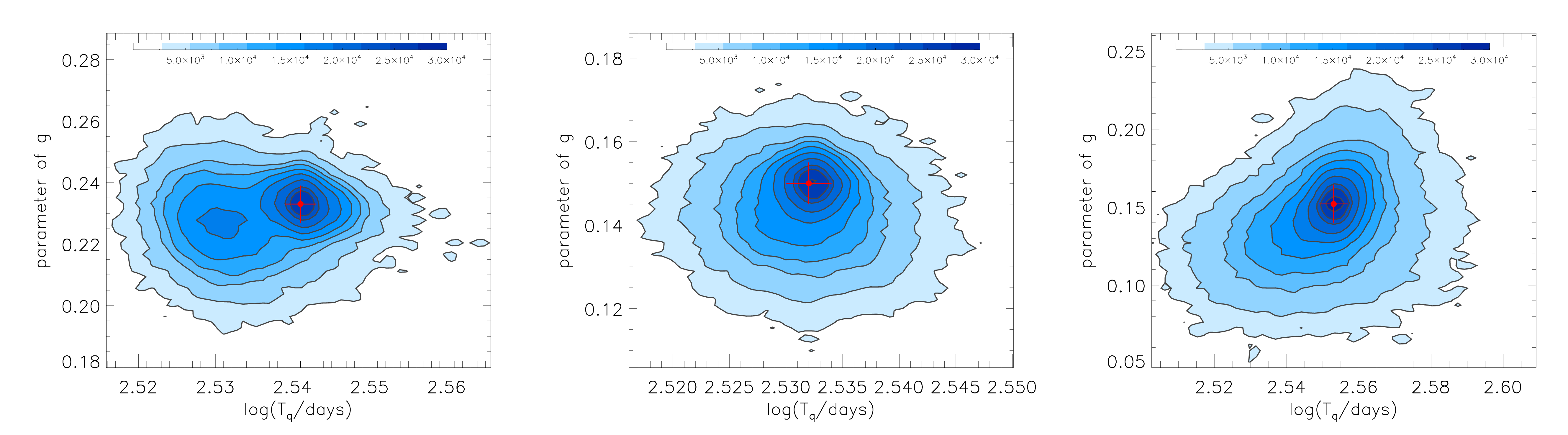}
\caption{The MCMC technique determined two-dimensional posterior distributions in contour of parameter $g$ and periodicity 
$\log(T_q)$ ($T_q$ in units of days) through the ZTF g-band (left panel) light curve, r-band (middle panel) light curve and i-band 
(right panel) light curve, respectively. In each panel, number densities related to different colors are shown in color bar, 
solid circle plus error bars in red show the accepted value and $1\sigma$ uncertainties of the parameters.}
\label{mcmc}
\end{figure*}

\subsection{Methods to determine optical QPOs in \obj}
%%%2

%%2.2
	Based on the high-quality long-term optical light curves from ZTF, the following four methods are applied to detect 
optical QPOs in \obj, similar as what have been done in \citet{pd20}.

\subsubsection{Direct fitting results by sine function}

%%2.2.1 - 1
	Based on a sine function plus a five-degree polynomial function applied to each ZTF light curve ($t$ in units of days 
as time information, $t_3$ as $t/1000$)
\begin{equation}
\begin{split}
LC_t~=~&a+b~\times~t_3+c~\times~t_3^2+d~\times~t_3^3+e~\times~t_3^4+f~\times~t_3^5\\
	&~+~g~\times~\sin(\frac{2\pi~t}{T_q}~+~\phi_0)
\end{split}
\end{equation},
the ZTF g/r/i-band light curves can be well simultaneously described through the maximum likelihood method combining with MCMC 
(Markov Chain Monte Carlo) \citep{fh13} technique with accepted well prior uniform distributions of the model parameters. Here, 
the only objective of applications of the model functions above is to check whether are there sine-like variability patterns 
(related to probable QPOs) included in the ZTF light curves, not to discuss physical origin of the sine-like variability patterns. 

\begin{table*}
\caption{Model parameters of Equation (1) leading to the best fitting results to ZTF light curves}
\begin{tabular}{cccccccccc}
\hline\hline
	&  $a$ & $b$ & $c$  & $d$ & $e$ & $f$ & $g$ & $\log(T_q)$ & $\phi_0$\\
\hline\hline
g-band  &  21.17$\pm$0.12  & -12.56$\pm$0.87 & 22.77$\pm$2.25 & -20.58$\pm$2.62 & 10.25$\pm$1.41 & 
	-2.15$\pm$0.28  & 0.233$\pm$0.006 & 2.541$\pm$0.002 & 1.16$\pm$0.87\\
\hline
r-band  &  20.03$\pm$0.07 & -8.60$\pm$0.51 & 14.96$\pm$1.31 & -12.25$\pm$1.53 & 5.26$\pm$0.83 &
	-0.96$\pm$0.16 & 0.150$\pm$0.005 & 2.532$\pm$0.002 & 0.85$\pm$0.05 \\
\hline
i-band &  21.44$\pm$0.58 & -25.73$\pm$4.99  &75.75$\pm$15.81 & -107.29$\pm$23.49 & 72.83$\pm$16.42 & -18.79$\pm$4.33 &
	0.152$\pm$0.012  & 2.553$\pm$0.004 & 1.90$\pm$0.21\\
\hline\hline
\end{tabular}\\
Notice: The first column shows which band ZTF light curve is considered. The ninth column shows the determined model parameter 
periodicity $\log(T_q)$ with $T_q$ in units of days.
\end{table*}

%%2.2.1 - 2
	Then, based on the MCMC technique determined posterior distributions of the model parameters, the accepted model 
parameters and corresponding $1\sigma$ uncertainties are listed in Table~1. Here, we do not show posterior distributions of all 
the model parameters, but Fig~\ref{mcmc} shows the MCMC technique determined two-dimensional posterior distributions of $g$ and 
periodicity $\log(T_q)$ of the ZTF g/r/i-band light curves. Meanwhile, based on the same functions, the Levenberg-Marquardt 
least-squares minimization technique (the known MPFIT package, \citealt{mpf09}) can lead to similar $1\sigma$ 
uncertainties of the model parameters, estimated through covariance matrix related to the determined model parameters, 
especially for $\log(T_q)$. Therefore, the determined $1\sigma$ uncertainties of $\log(T_q)$ are reliable enough in \obj. 
Moreover, in order to show more clearer sine variability patterns, the determined sine-like component is shown in dashed 
line in each ZTF light curve in top left panel of Fig.~\ref{lmc}.

%%2.2.1 - 3
	Based on the determined model parameters, left panels of Fig.~\ref{lmc} show the best fitting results and corresponding 
residuals (light curve minus the best fitting results) to each ZTF band light curve, leading $\chi^2/dof$ (dof=885 as degree of 
freedom) to be 5.76. Meanwhile, after removing the five-degree polynomial component in each band light curve, accepted the 
determined periodicities $T_q$, corresponding phase-folded light curves can also be well described by sine function 
$\sin(2\pi~ph+\phi_0)$ with $ph$ as phase information. The best fitting results and corresponding residuals to the folded ZTF 
g/r/i-band light curves are shown in right panels of Fig.~\ref{lmc}, leading to $\chi^2/dof\sim5.89$.

%%2.2.1 - 4
	Based on the results in Fig.~\ref{lmc}, there are apparent optical QPOs with periodicities about $348\pm2$ days, 
$340\pm2$ days, $357\pm4$ days in the ZTF g/r/i-band light curves, respectively. The totally similar periodicities in the ZTF 
g/r/i-band light curves provide strong evidence to support the optical QPOs in \obj. Moreover, considering shorter time 
duration of the ZTF i-band light curve, a bit difference between the periodicity in i-band light curve and the periodicities 
in g/r-band light curves can be accepted.

\begin{figure}
\centering\includegraphics[width = 8cm,height=5cm]{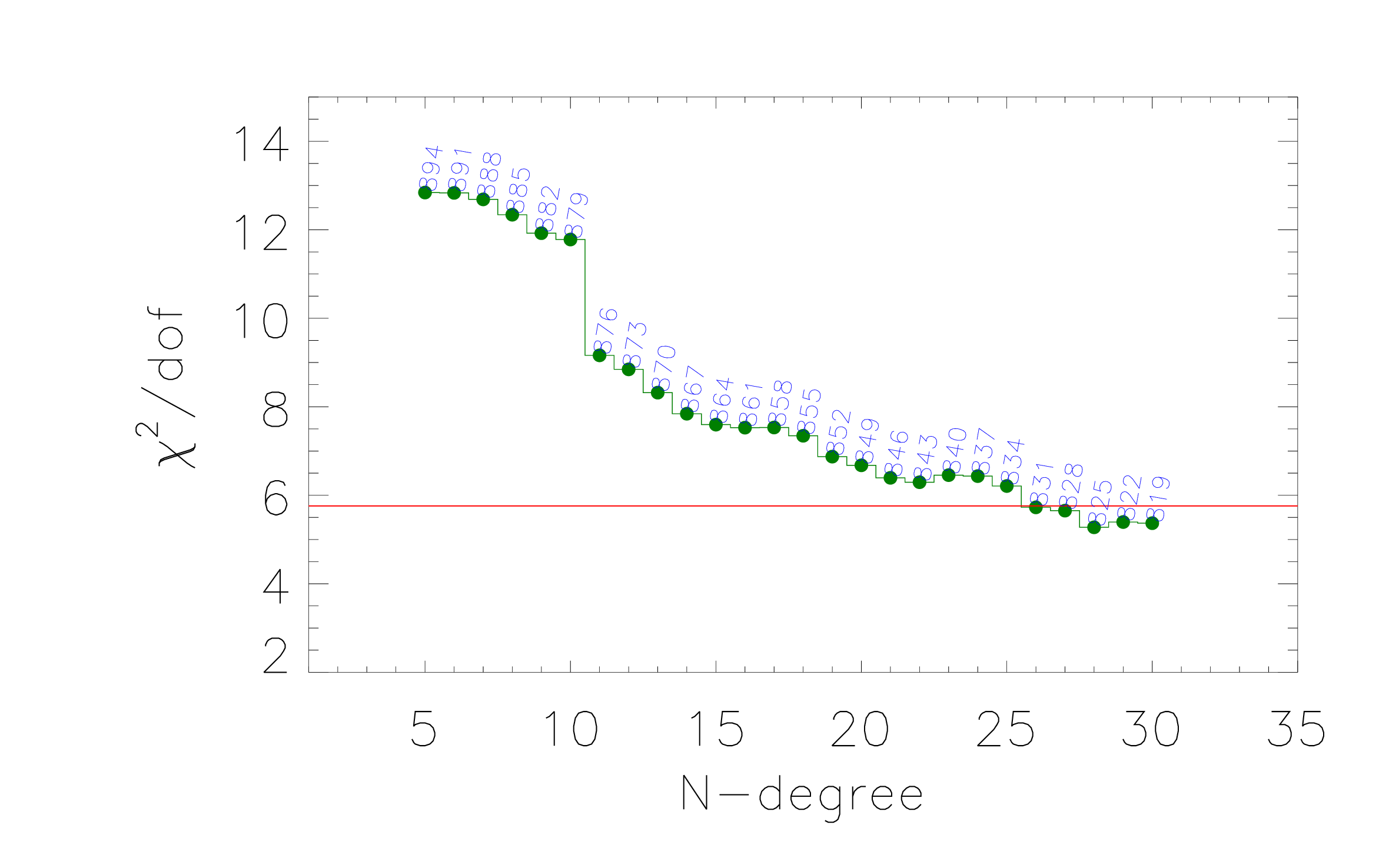}
\caption{Dependence of $\chi2/dof$ on degree of applied N-degree of polynomial function to describe ZTF light curves. Horizon 
red line marks $\chi2/dof=5.76$ (the value by the best descriptions to ZTF light curves by model function in Equation (1)). 
Blue character related to each data point marks corresponding value of dof.}
\label{cdf}
\end{figure}

%%2.2.1 - 5
	In order to confirm the apparent optical QPOs in ZTF light curves, rather than the model function shown in Equation (1), 
a N-degree (N=5, \dots, 30) polynomial function without any restrictions to model parameters is applied to re-describe each ZTF 
light curve. Fig.~\ref{cdf} shows dependence of re-determined $\chi^2/dof$ by polynomial function on the degree of the polynomial 
function. It is clear that N$\ge$26 can also lead to well accepted descriptions with $\chi^2/dof\sim4759.44/831=5.73\le5.76$ to 
ZTF light curves of \obj. Then, similar as what we have recently done in \citet{zh22b}, through determined different $\chi^2/dof$ 
for model function (M1) in Equation (1) and for the 26-degree polynomial function (M2) applied, the calculated $F_p$ value is about
\begin{equation}
	F_p=\frac{\frac{\chi^2_{M1}-\chi^2_{M2}}{dof_{M1}-dof_{M2}}}{\chi^2_{M2}/dof_{M2}}\sim9.7\times10^{-5}
\end{equation}
Based on $dof_{M1}-dof_{M2}$ and $dof_{M2}$ as number of dofs of the F distribution numerator and denominator, the expected 
confidence level is quite smaller than $10^{-7}$ to support the 26-degree polynomial function. In other words, confidence level 
quite higher than 6$\sigma$ to support that the model function in Equation (1) is more preferred.

%%2.2.1 - 6
	Therefore, based on the direct fitting results by sine function, there are apparent and reliable optical QPOs with 
periodicities about $348\pm2$ days, $340\pm2$ days, $357\pm4$ days in the ZTF g/r/i-band light curves, respectively. And, 
through the F-test technique, the sine component included in the ZTF light curves of \obj~ is more preferred with confidence 
level higher than 6$\sigma$.

%%%3

\subsubsection{Results from Generalized Lomb-Scargle (GLS) periodogram}

\begin{figure}
\centering\includegraphics[width = 8cm,height=9cm]{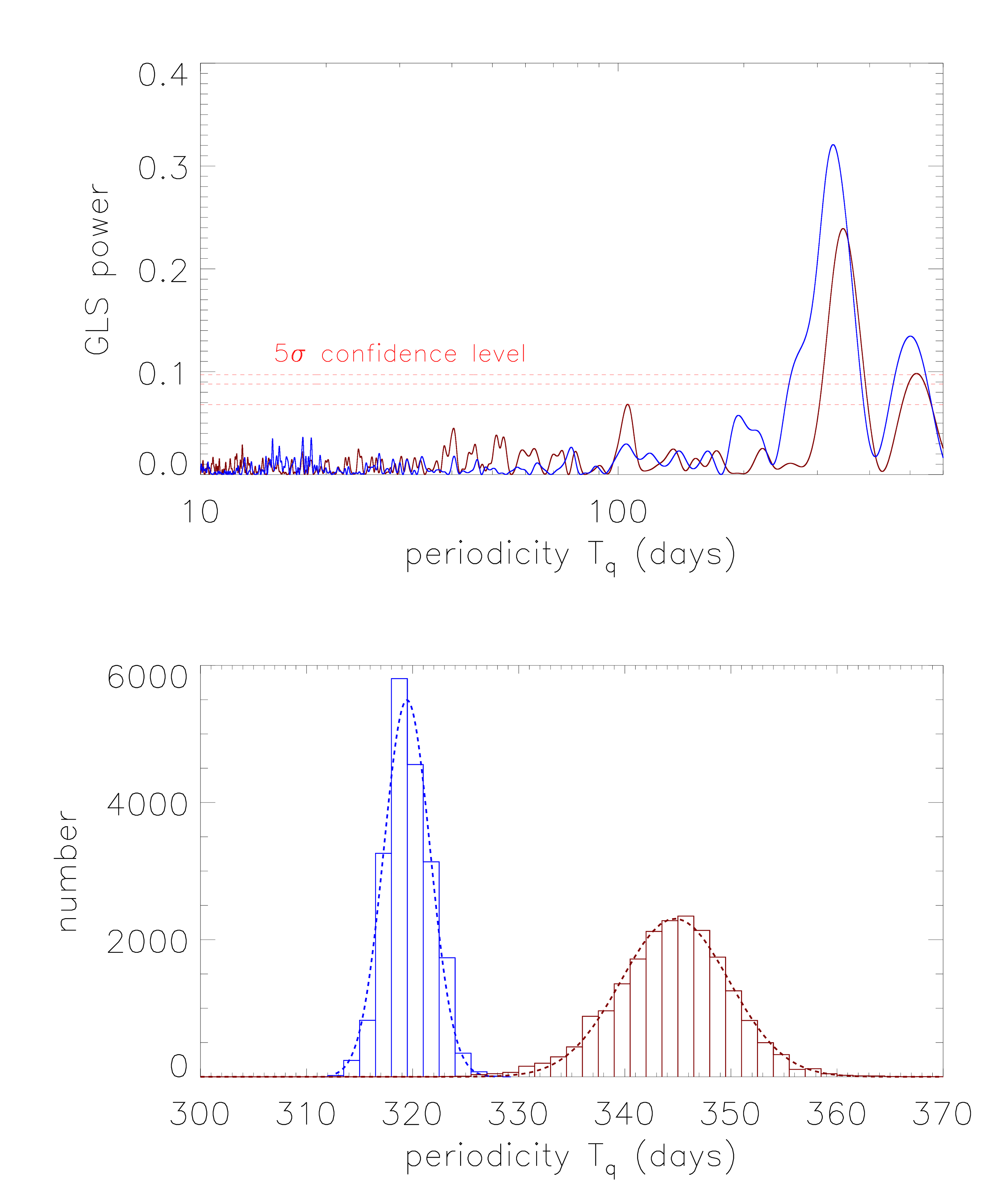}
\caption{Top panel shows power properties through the Generalized Lomb-Scargle periodogram applied to g/r-band light curves. 
Solid line in blue and in purple show the results through the g-band and r-band light curve, respectively. From top to bottom, 
horizontal dashed red lines show the $5\sigma$, $4\sigma$ and $3\sigma$ confidence levels through the bootstrap method 
(false-alarm probabilities of 5.3e-7, 6.3e-5 and 2.7e-3). Bottom panel shows distributions of GLS periodogram determined 
peak positions considering 20000 re-build light curves through the ZTF g/r-band light curves. In bottom panel, histogram 
filled by blue lines and filled by purple lines show corresponding results through the g-band and r-band light cures, 
respectively, and thick dashed line in the same color shows corresponding Gaussian described results to the distribution.
}
\label{gls}
\end{figure}

%%2.2.2 - 1
	In order to provide further evidence to support the optical QPOs in \obj, besides the direct fitting results by sine 
function in Fig.~\ref{lmc}, the widely accepted Generalized Lomb-Scargle (GLS) periodogram \citep{ln76, sj82, zk09, vj18} 
(included in the python package of astroML.time\_series) is applied to check the periodicities in the observed ZTF g/r-band 
light curves in \obj. Here, due to small number of data points and short time duration of the ZTF i-band light curve, the GLS 
periodogram is not applied to the i-band light curve. Top panel of Fig.~\ref{gls} shows the GLS power properties. It is clear 
that there is one periodicity around 330 days with confidence level higher than $5\sigma$ (the false-alarm probability of 5e-7) 
determined by the bootstrap method as discussed in \citet{ic19}.

%%2.2.2 - 2
	Moreover, in order to determine uncertainties of GLS periodogram determined periodicity, the well-known bootstrap method 
is applied as follows. Through the observed ZTF g/r-band light curves, more than half of data points are randomly collected to 
re-build a new light curve. Then, within 20000 rebuild light curves after 20000 loops, the same GLS power properties are applied 
to determine new periodicities related to the rebuild light curves. Bottom panel of Fig.~\ref{gls} shows distributions of 
corresponding 20000 GLS periodogram determined periodicities related to the 20000 rebuild light curves. And through the 
Gaussian-like distributions, the determined periodicities and corresponding $1\sigma$ uncertainties are 319$\pm$2 days and 
344$\pm$5 days in the rebuild light curves through the g-band and r-band light curves, respectively, strongly reconfirming the 
smaller uncertainties of the periodicities determined by the MCMC technique. Moreover, through properties of g-band light curve, 
the tiny difference in the periodicity determined through different methods could be due to probably large uncertainties in 
g-band light curves leading the applied polynomial component to be not so appropriate. The GLS periodogram determined 
periodicities are consistent with results shown in Fig.~\ref{lmc}, to confirm the optical QPOs in \obj.

\begin{figure}
\centering\includegraphics[width = 8cm,height=9cm]{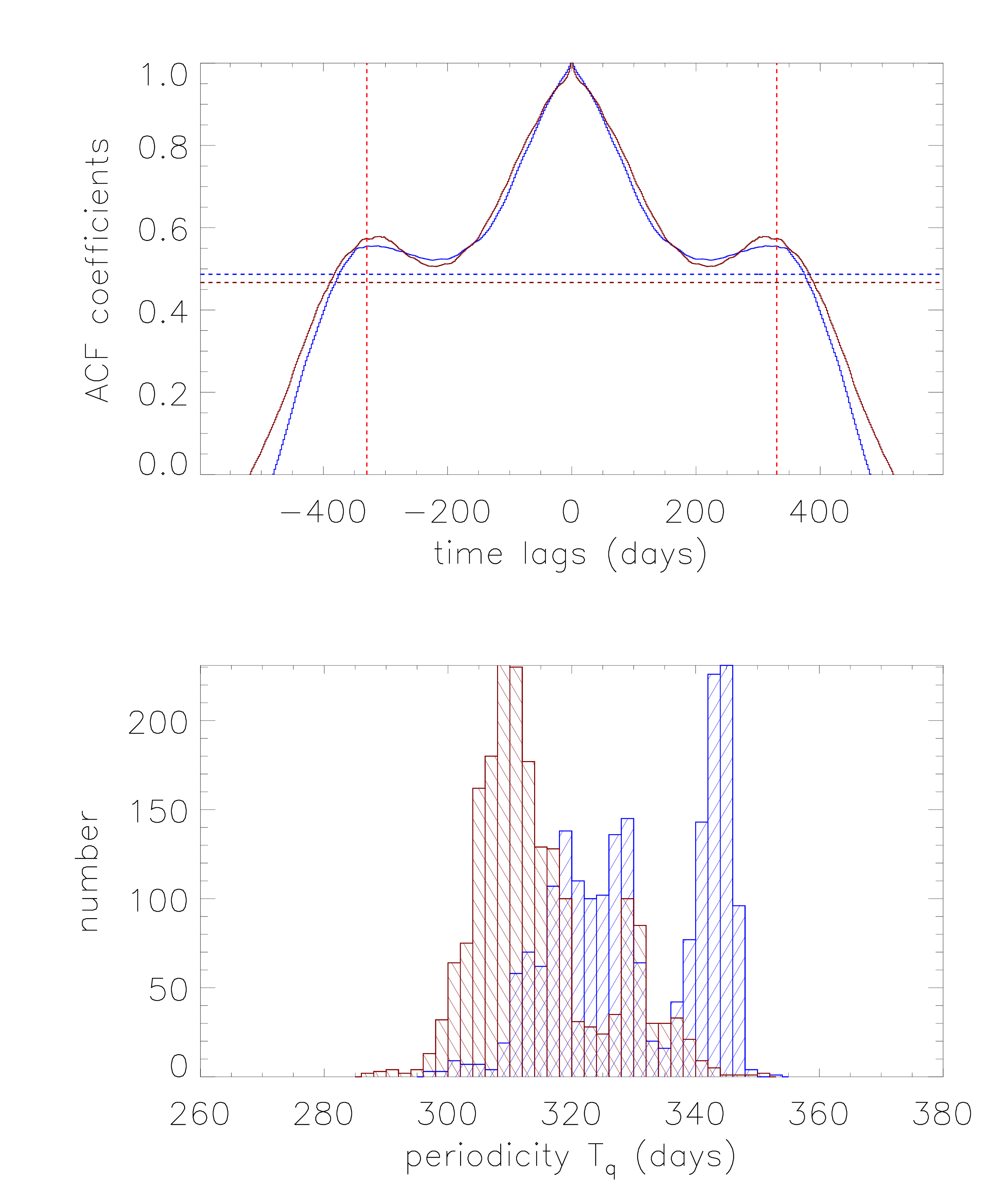}
\caption{Top panel shows properties of the ACF coefficients. Solid blue line and solid purple line represent the results 
through the ZTF g-band and r-band light curves, respectively, vertical dashed red lines mark positions with $T_q\sim\pm340$ 
days. Horizontal dashed blue line and horizontal purple dashed line show the Monte Carlo method determined 5$\sigma$ confidence 
level for the coefficient at time lags around $\pm$340 days for the ZTF g-band and r-band light curves, respectively. Bottom 
panel shows distributions of the bootstrap method determined 2000 periodicities. In bottom panel, histogram fill by blue 
lines and histogram filled by purple lines show the results through the ZTF g-band and r-band light curve, respectively.}
\label{acf}
\end{figure}

\subsubsection{Results through the Auto-cross Correlation Function}

%%2.2.3 - 1
	Moreover, similar as what we have recently done on optical QPOs in \citet{zh22a, zh22c}, the Auto-cross Correlation 
Function (ACF) is applied to check optical QPOs in observed ZTF g/r-band light curves of \obj. Corresponding results are 
shown in top panel of Fig.~\ref{acf}, totally similar periodicities around 340 days can be confirmed, to support the optical 
QPOs in \obj. Here, direct linear interpolation is applied to the ZTF light curves, leading to evenly sampled light curves, 
and then the IDL procedure djs\_correlate.pro (written by David Schlegel, Princeton) is applied to determine the correlation 
coefficients at different time lags.

%%2.2.3 - 2
	Furthermore, the common Monte Carlo method is applied as follows to determine confidence level for ACF results. 
Accepted the time information of the observed ZTF g/r-band light curves of \obj, 3.2 million light curves are randomly 
created by white noise process. For the $i$th randomly created light curve of white noise, similar procedure is applied to 
determine the correlation coefficients, to determine the maximum coefficient $Coe_{i}$ (i=1, \dots, $3.2\times10^6$) with 
time lags between 200 days and 500 days. Then, among all the 3.2 million values of $Coe$, the maximum value of 0.4866 
(0.46686) is determined as corresponding value for the 5$\sigma$ confidence level (probability of $\frac{1}{3.2\times10^{6}}$) 
for the ACF results through the ZTF r-band (g-band) light curve. Therefore, the determined confidence level is higher than 
5$\sigma$ for the ACF determined optical QPOs in \obj.

%%2.2.3 - 3
	Here, as well known that 5$\sigma$ corresponds to a probability of $3\times10^{-7}$ corresponding to about 1 in 3.2 
million, therefore, 3.2 million light curves are created by white noise process, in order to determine 5$\sigma$ confidence 
level for ACF results (and also for the following WWZ results in the subsection 2.2.4). Furthermore, due to the following two 
main reasons, applications of red noise time series are not appropriate to determine 5$\sigma$ confidence levels of ACF 
results (and for the following WWZ results). On the one hand, ACF results for red noise time series can be described by an 
exponential function depending on intrinsic time intervals and correlation coefficients between adjacent data points. On the 
other hand, as more recent discussions in \citet{km21}, there are detections of fake periodic signals in red noise time series. 
Therefore, rather than red noise time series, white noise time series are preferred to be applied to determine confidence levels 
of ACF results (and the following WWZ results).

%%2.2.3 - 4
	Meanwhile, similar bootstrap method is applied to determine uncertainties of ACF method determined periodicities in \obj. 
Through the observed ZTF g/r-band light curves, more than half of data points are randomly collected to re-build a new light 
curve. Then, within 2000 rebuild light curves after 2000 loops, the ACF method is applied to determine new periodicities related 
to the rebuild light curves. Bottom panel of Fig.~\ref{acf} shows distributions of corresponding 2000 ACF method determined 
periodicities related to the 2000 rebuild light curves. However, due to unevenly sampled ZTF light curves and applications 
linear interpolation, distributions in bottom panel of Fig.~\ref{acf} are not well Gaussian-like, but standard deviations about 
12 days and 10 days of the distributions can be safely accepted as uncertainties of ACF method determined periodicities through 
ZTF g-band and r-band light curves.

\subsubsection{Results through the WWZ method}

\begin{figure}
\centering\includegraphics[width = 8cm,height=14cm]{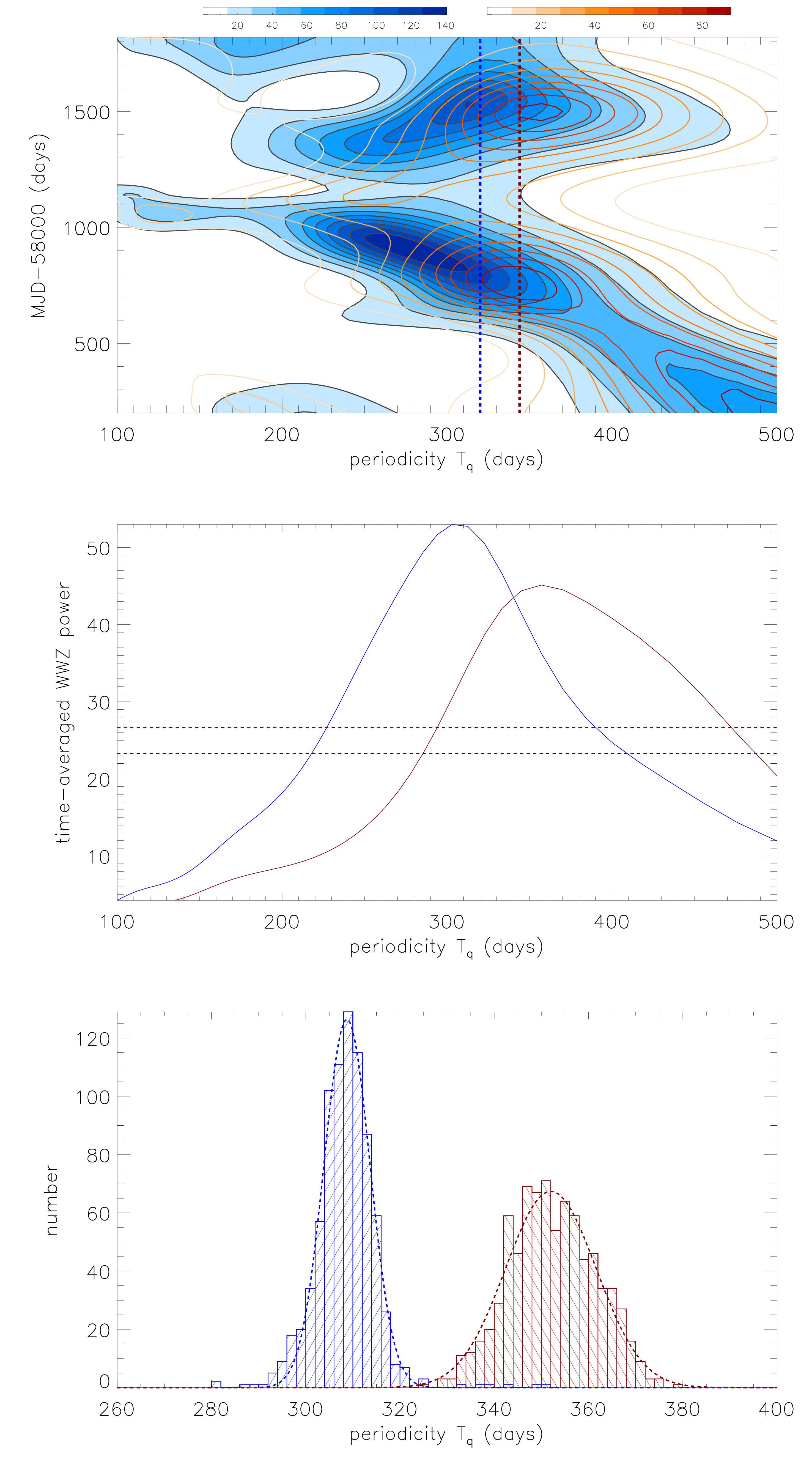}
\caption{Top panel shows the two dimensional power maps determined by the WWZ method with frequency step of 0.0001 and searching 
periodicities from 100 days to 500 days applied to the ZTF g/r-band light curves. In top panel, vertical dashed lines mark WWZ 
method determined periodicities. Contour filled with bluish colors represent the results through the ZTF g-band light curve, 
contour with levels shown in reddish colors represent the results through the ZTF r-band light curve. In top regions of top panel, 
color bars show corresponding number densities for contour levels in different colors. Middle panel shows the WWZ method determined 
time-averaged power properties. Solid blue line and solid purple line show the results through the ZTF g-band and r-band light 
curves, respectively, and horizontal dashed blue line and horizontal dashed purple line mark the corresponding 5$\sigma$ confidence 
levels. Bottom panel shows the bootstrap method determined periodicity distributions. In bottom panel, histogram filled by blue 
lines and filled by purple lines show corresponding results through the g-band and r-band light cures, respectively, and thick 
dashed line in the same color shows corresponding Gaussian described results to the distribution.}
\label{wwz}
\end{figure}

%%2.2.4 - 1
        Moreover, similar as what we have recently done on optical QPOs in \citet{zh22a, zh22c}, the WWZ method \citep{fg96, 
al16, gt18, ks20,ly21} is also applied to check the optical periodicities in the observed ZTF g/r-band light curves of \obj. 
Corresponding results on both two dimensional power map properties and time-averaged power properties are shown in top and 
middle panels of Fig.~\ref{wwz}, totally similar periodicities around 340 days can be confirmed, to support the optical QPOs 
in \obj. Here, the python code wwz.py written by M. Emre Aydin is applied in the manuscript.

%%2.2.4 - 2
	Meanwhile, the similar common Monte Carlo method is applied to determine confidence level for the WWZ determined 
results. Among the 3.2 million randomly created light curves for white noises, for the $i$th randomly created light curve 
of white noise, maximum value $Mp_i$ of the WWZ method determined time-averaged power spectra can be well determined. Then, 
among all the 3.2 million values of $Mp$, the maximum value of 26.64 (21.68) is determined as corresponding value for the 
5$\sigma$ confidence level for the WWZ method determined time-averaged power properties through the ZTF g-band (r-band) 
light curve, which are shown in middle panel of Fig.~\ref{wwz}. Therefore, the determined confidence level is higher than 
5$\sigma$ for the WWZ method  determined QPOs in \obj.

%%2.2.4 - 3
	Furthermore, the similar bootstrap method is applied to determine uncertainties of the WWZ method determined 
periodicities in \obj. Through the observed ZTF g/r-band light curves, more than half of data points are randomly collected 
to re-build a new light curve. Then, within 800 rebuild light curves after 800 loops, the same WWZ method determined 
time-averaged power properties are applied to measure the new periodicities related to the rebuild light curves. Bottom 
panel of Fig.~\ref{wwz} shows distributions of corresponding WWZ method determined periodicities related to the 800 rebuild 
light curves. Through the Gaussian-like distributions, the determined periodicities and corresponding $1\sigma$ uncertainties 
are 308$\pm$5 days and 352$\pm$10 days in the rebuild light curves through the g-band and r-band light curves, respectively, 
to re-confirm the reliable optical QPOs in \obj.

\begin{table}
\caption{Periodicities of optical QPOs by different methods in \obj}
\begin{tabular}{cccc}
\hline\hline
Method & band & $\log(T_q)$ & CL\\
\hline\hline
	\multirow{3}{*}{DF} & g-band  &  348$\pm$2   &  \multirow{3}{*}{$>6\sigma$}\\  
		   & r-band  &   340$\pm$2     &    \\
		   & i-band  &   357$\pm$4     &    \\
\hline
	\multirow{2}{*}{GLS} & g-band  &   319$\pm$2  & \multirow{2}{*}{$>5\sigma$}\\
                    & r-band  &  344$\pm$5   &  \\
\hline
	\multirow{2}{*}{ACF} & g-band  & 327$\pm$12    & \multirow{2}{*}{$>5\sigma$}\\
	 & r-band  &  309$\pm$10   &  \\
\hline
	\multirow{2}{*}{WWZ} & g-band  &  308$\pm$5   & \multirow{2}{*}{$>5\sigma$}\\
         & r-band  &  352$\pm$10   &  \\
\hline\hline
\end{tabular}\\
Notice: The first column shows which method is applied to determine optical QPOs in \obj, DF means the 'direct fitting results 
to ZTF light curves' as discussed in subsection 2.2.1. The second column shows which ZTF band light is considered. The third 
column shows the determined periodicity $T_q$ in units of days, the last column shows the determined corresponding confidence level. 
\end{table}

\subsubsection{Conclusion on the robustness of the optical QPOs in \obj}

%%2.2.5 - 1
	Based on the different methods discussed above to detect reliable optical QPOs, Table~2 lists the necessary information 
of the periodicities and confidence levels. Therefore, the robust optical QPOs with periodicities around 340 days (0.93 years) 
in \obj~ can be detected and well confirmed from the 4.45 years-long ZTF light curves (time duration about 4.7 times longer than 
the detected periodicities) with confidence level higher than $5\sigma$, based on the best-fitting results directly by the sine 
function shown in the left panels of Fig.~\ref{lmc}, on the sine-like phase-folded light curve shown in the right panels of 
Fig.~\ref{lmc}, on the results of GLS periodogram shown in Fig.~\ref{gls}, and on properties of ACF coefficients shown in 
Fig.~\ref{acf} and on the power properties determined by the WWZ method shown in Fig.~\ref{wwz}. The results cab be well applied 
to guarantee the robustness of the optical QPOs in \obj.

%%2.2.5 - 2
	Based on the well determined reliable and robust optical QPOs in \obj, we can find that the periodicity around 340 days 
in \obj~ is so-far the smallest periodicity among the reported long-standing optical QPOs in normal broad line AGN in the 
literature as described in Introduction. The results are strongly indicating that BBH systems with shorter periodicities could 
be well detected through the ZTF sky survey project. More interestingly, to detect more optical QPOs related to BBH systems 
with shorter periodicities through ZTF light curves could provide more stronger BBH candidates for expected background 
gravitational wave signals at nano-Hz frequencies.

\begin{figure*}
\centering\includegraphics[width = 18cm,height=5.5cm]{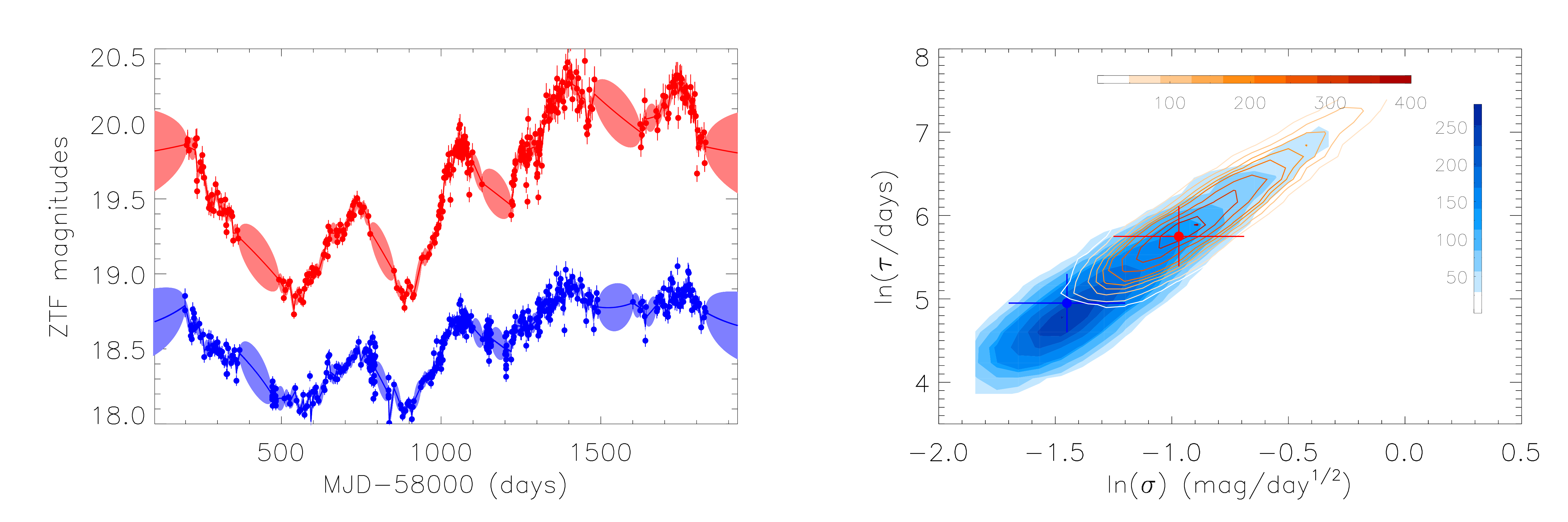}
\caption{Left panel shows the JAVELIN code determined best descriptions to the long-term ZTF g-band (solid circles plus 
error bars in red) and r-band (solid circles plus error bars in blue) light curves of \obj. Solid line and area filled 
in blue and in red show the best descriptions and corresponding $1\sigma$ confidence bands to the g-band light curve and 
to the r-band light curve, respectively. Right panel shows the MCMC technique determined two-dimensional posterior 
distributions in contour of $\ln(\tau)$ ($\tau$ in units of days) and $\ln(\sigma)$ ($\sigma$ in units of $mag/day^{1/2}$). 
Contour filled with bluish color represents the results through the r-band light curve, and contour with level in reddish 
color shows the results through the g-band light curve. In right panel, solid circle plus error bars in blue and in red 
show the accepted values and $1\sigma$ uncertainties of $\ln(\tau)$ and $\ln(\sigma)$ to the r-band light curve and to 
the g-band light curve, respectively.
}
\label{drw}
\end{figure*}

%%%%should be corrected by F-test applied to determined 999999\% confidence bands
\begin{figure*}
\centering\includegraphics[width = 18cm,height=5.5cm]{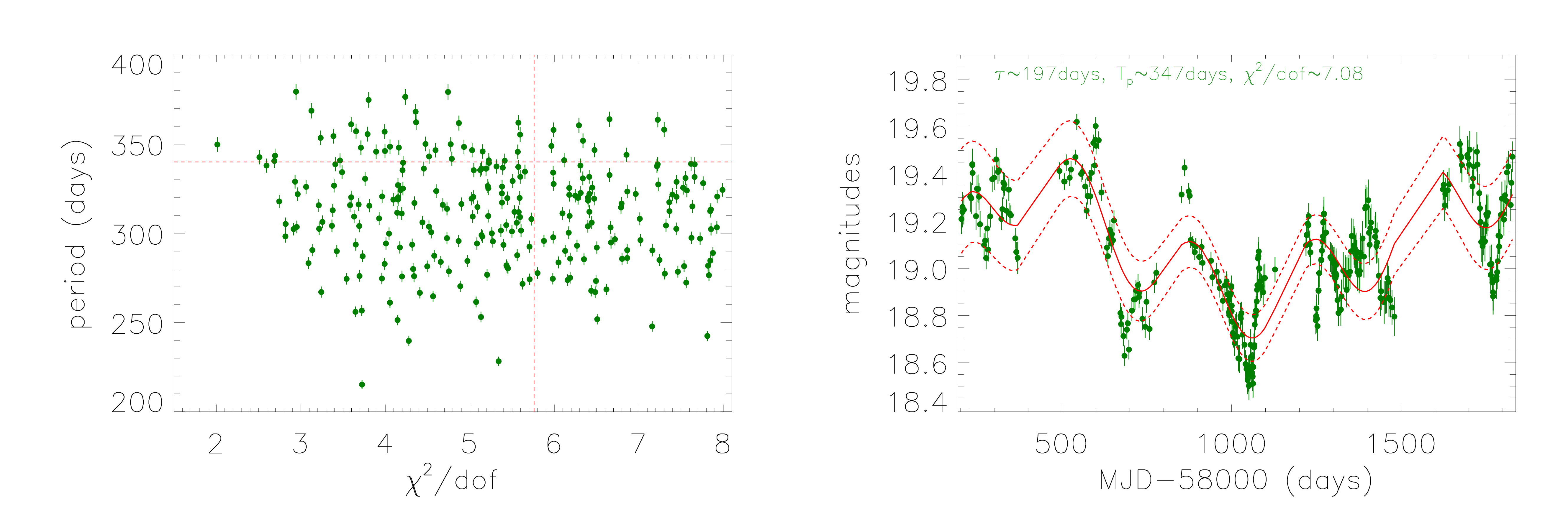}
\caption{Left panel shows properties of $\chi^2/dof$ and determined periodicity of the 263 CAR process simulated light 
curves with probably mis-detected QPOs. In left panel, horizontal red dashed line marks the position of periodicity 344 days, 
the periodicity from r-band light curve of \obj, vertical red dashed line marks the position of $\chi^2/dof\sim5.76$, 
the value for r-band light curve of \obj. Right panel shows an example on probably mis-detected QPOs in the simulating 
light curves by the CAR process. In right panel, solid dark green circles plus error bars show the simulated light curve, 
solid and dashed red lines show the best descriptions and corresponding $5\sigma$ confidence bands to the light curve, 
based on a sine function plus a five-degree polynomial function. For the shown simulated light curve, the input parameter 
$\tau$, the determined periodicity $T_p$ and $\chi^2/dof$ are listed and shown in characters in dark green in top corner 
in right panel.
}
\label{fake}
\end{figure*}

\section{Main Discussions}
\subsection{Mis-detected optical QPOs related to central intrinsic AGN activities in \obj?}

%%3.1 - 1
	Due to short time durations of ZTF light curves, it is necessary and interesting to check whether the determined 
optical QPOs was mis-detected QPOs tightly related to central intrinsic AGN activities of \obj, although four different 
mathematical methods applied in the section above can provide robust evidence to support the detected optical QPOs in \obj. 
Similar as what we have recently done in \citet{zh22a, zh22c} to check probability of mis-detected QPOs in AGN activities 
in SDSS J0752 and in SDSS J1321, the following procedure is applied.

%%%3.1 - 2
	As well discussed in \citet{kbs09, koz10, mi10, zk13, zk16, zh17, tm18, mv19, sr22}, the known Continuous AutoRegressive 
(CAR) process and/or the improved Damped Random Walk (DRW) process can be applied to describe the fundamental AGN activities 
\citep{mr84, um97, ms16, bg20, bs21}. Here, based on the DRW process, the public code JAVELIN (Just Another Vehicle for 
Estimating Lags In Nuclei) \citep{koz10, zk13} is firstly applied to describe the ZTF r-band light curve, with two process 
parameters of intrinsic characteristic variability amplitude and timescale of $\sigma$ and $\tau$. The best descriptions are 
shown in left panel of Fig.~\ref{drw}. And corresponding MCMC determined two dimensional posterior distributions of $\sigma$ 
and $\tau$ are shown in right panel of Fig.~\ref{drw}, with the determined $\ln(\tau/days)\sim4.95\pm0.35$ 
($\tau\sim140_{-40}^{+60}$ days) and $\ln(\sigma/(mag/dyas^{1/2}))\sim-1.45\pm0.25$. Here, because of the totally same 
periodicities through the ZTF r-band light curve by different methods in the section above, the r-band rather than the g-band 
light curve is considered in the section. Certainly, Fig.~\ref{drw} also shows corresponding results through the g-band light 
curve.

%%%3.1 - 3
	Then, probability of mis-detected QPOs from DRW process described intrinsic AGN variabilities can be estimated as 
follows, through applications of the CAR process discussed in \citet{kbs09}:
\begin{equation}
\dif LC_t=\frac{-1}{\tau}LC_t\dif t+\sigma_*\sqrt{\dif t}\epsilon(t)~+~19.21
\end{equation}
with $\epsilon(t)$ as a white noise process with zero mean and variance equal to 1. Here, the value 19.21, the mean value 
of the ZTF r-band light curve of \obj, is set to be the mean value of $LC_t$. Then, a series of 100000 simulating light curves 
[$t$,~$LC_t$] are created, with $\tau$ randomly selected from 100 days to 200 days (range of the determined $\tau$ of \obj, 
after considering uncertainties). Then, variance $\tau\sigma_*^2/2$ of CAR process created light curve is set to be 0.168 
which is the variance of ZTF r-band light curve of \obj. And, time information $t$ are the same as the observational time 
information of ZTF r-band light curve of \obj. And similar uncertainties $LC_{t}\times\frac{LC_{err}}{LC_{r}}$ are simply added 
to the simulating light curves $LC_t$, with $LC_{r}$ and $LC_{err}$ as the ZTF r-band light curve and corresponding 
uncertainties of \obj.

%%%3.1 - 4
	Among the 100000 simulated light curves, there are 263 light curves collected with reliable mathematical determined 
periodicities according to the following four simple criteria. First, the simulated light curve can be well described by 
equation (1) with corresponding $\chi^2/dof$ smaller than 8 ($\chi^2/dof\sim5.8$ in Fig.~\ref{lmc}). Second, the simulated 
light curve has apparent GLS periodogram determined peak with corresponding periodicity smaller than 500 days with confidence 
level higher than $5\sigma$. Third, the simulated light curve has apparent ACF and WWZ method determined peak with corresponding 
periodicity smaller than 500 days with confidence level higher than $5\sigma$. Fourth, the determined periodicity by equation 
(1) is consistent with the GLS periodogram determined periodicity (which are similar as the ACF and WWZ method determined 
periodicities) within 10 times of the determined uncertainties by applications of equation (1). Left panel of Fig.~\ref{fake} 
shows properties of the determined $\chi^2/dof$ related to the best fitting results determined by equation (1) and the GLS 
periodogram determined periodicities (which are similar as the ACF and WWZ method determined periodicities) of the 263 
simulated light curves. And right panel of Fig.~\ref{fake} shows one of the 263 light curves with best fitting results by 
equation (1). 

%%%3.1 - 5
	Therefore, even without any further considerations, it can be confirmed that the probability is only 0.26\% 
(263/100000) (confidence level higher than $3\sigma$) to support mis-detected QPOs in CAR process simulated light curves 
related to AGN activities. Furthermore, accepted uncertainty $\Delta_T=5$ days of optical periodicity in ZTF r-band light 
curve, if to limit periodicities within range from $344\pm5\times\Delta_T$ in the simulated light curves, there are only 
112 simulated light curves collected, leading to the probability only about 0.11\% (112/100000) (confidence level higher 
than $3.2\sigma$) to support mis-detected QPOs in AGN activities. In other words, the confidence level higher than $3\sigma$ 
to confirm the optical QPOs not from intrinsic AGN activities in \obj, although short time durations in ZTF light curves, 
after well considering effects of AGN activities described by CAR process.

\begin{figure}
\centering\includegraphics[width = 8cm,height=5cm]{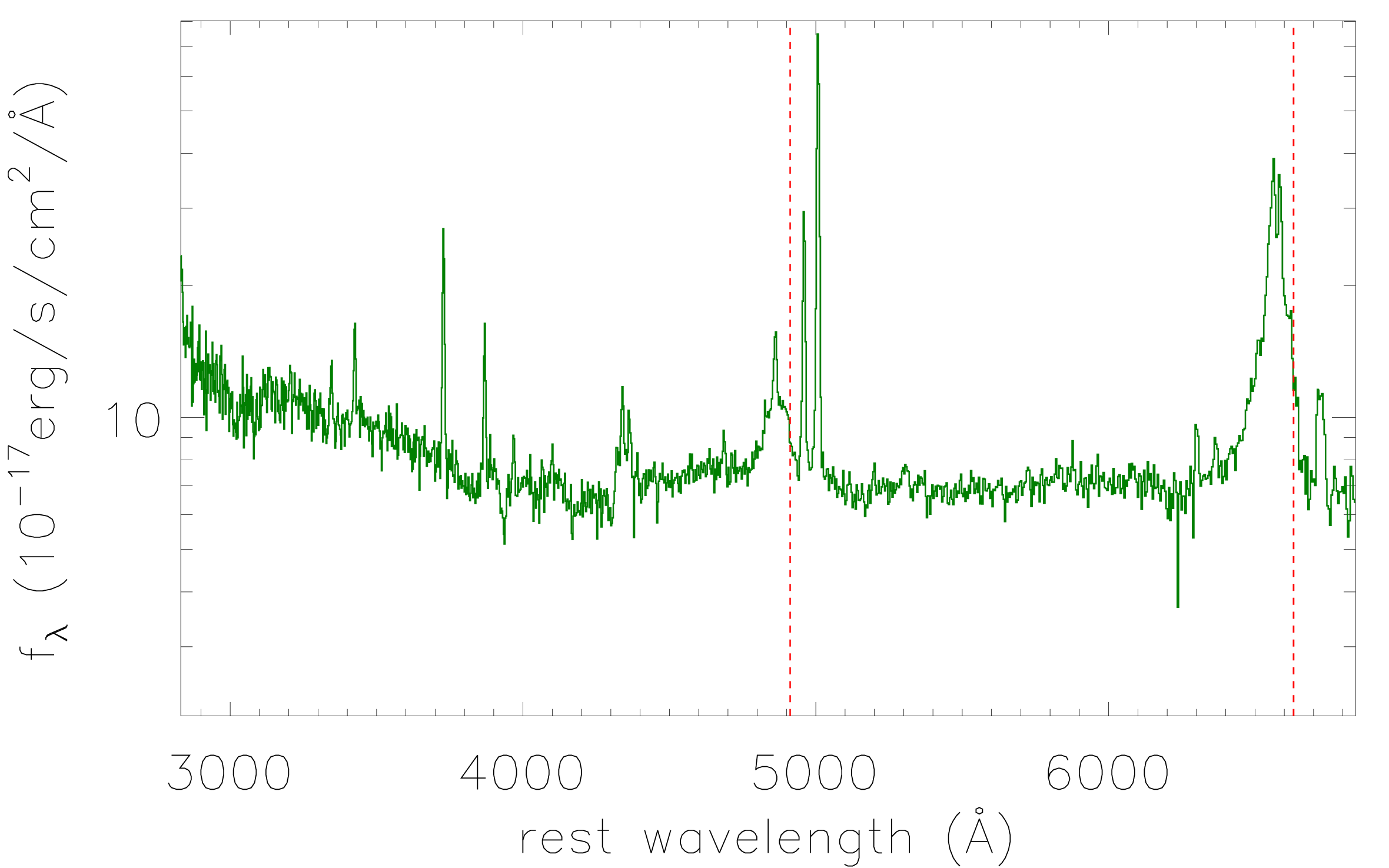}
\caption{SDSS spectrum of \obj~ in rest frame. Vertical dashed red lines mark the shoulders in broad Balmer emission lines.
}
\label{spec}
\end{figure}

\begin{figure*}
\centering\includegraphics[width = 8cm,height=6cm]{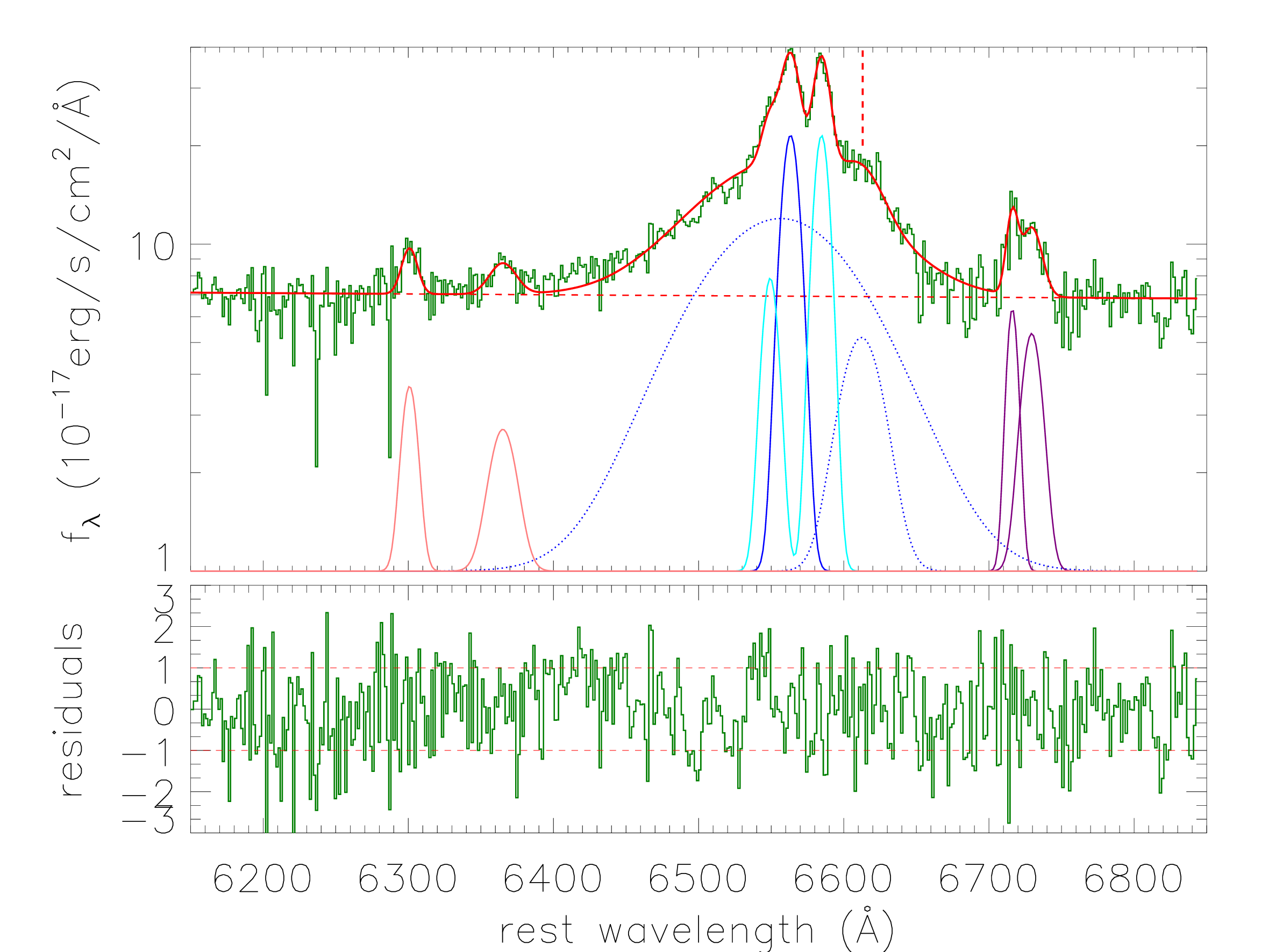}
\centering\includegraphics[width = 8cm,height=6cm]{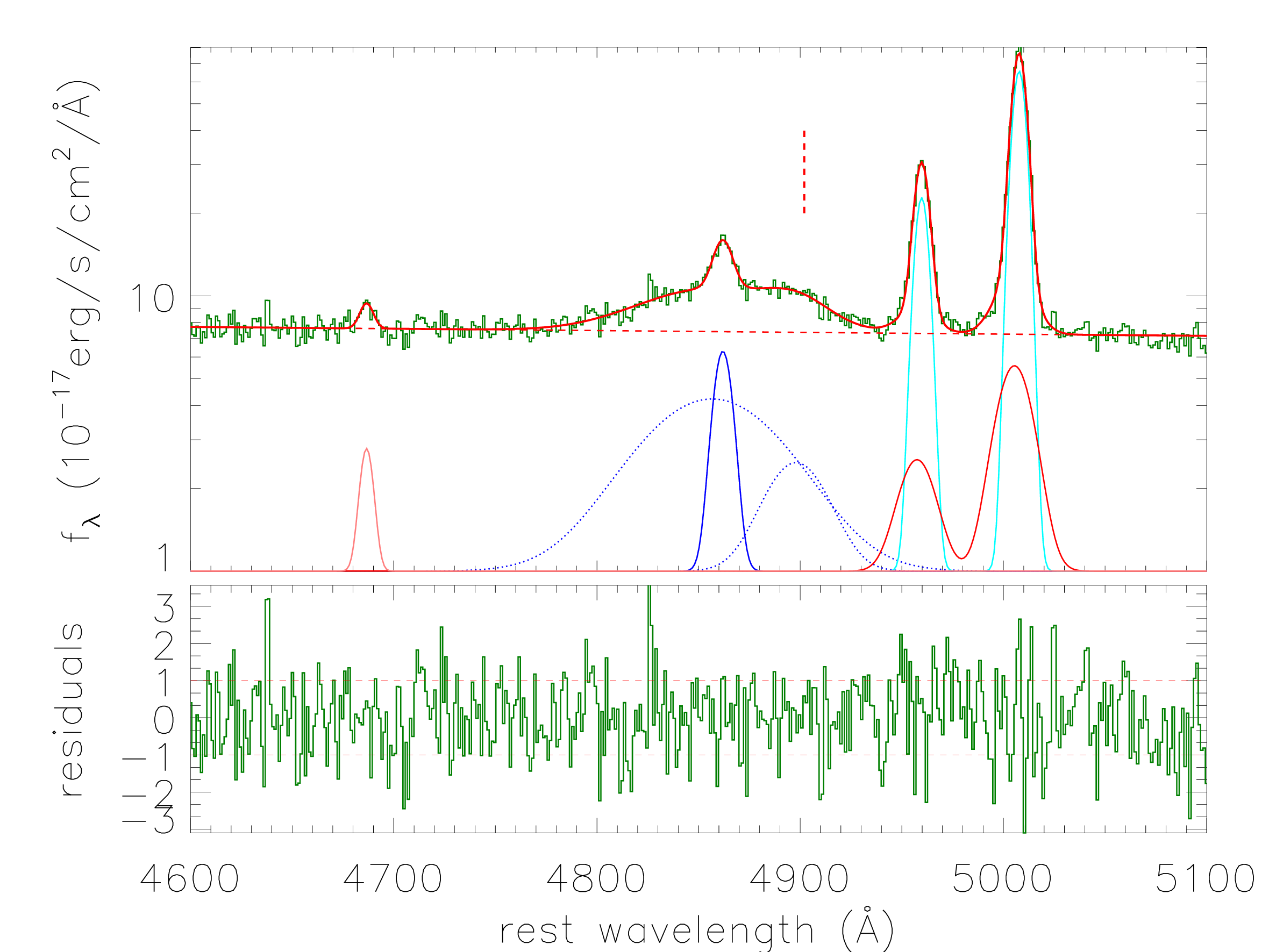}
\caption{Left panels show the best-fitting results (top panel) and corresponding residuals (bottom panel) (line spectrum 
minus the best fitting results and then divided by uncertainties of the line spectrum) to the emission lines around H$\alpha$. 
Right panels show the results to the emission lines around H$\beta$. In each top panel, solid dark green line shows the SDSS 
spectrum, solid red line shows the best fitting results, dashed blue lines show the determined two broad Gaussian components 
in broad Balmer line, solid blue line shows the determined narrow Gaussian component in narrow Balmer line, dashed red line 
shows the determined power law continuum emissions. In top right panel, solid cyan and solid dark red lines show the determined 
core components and components related to wings in [O~{\sc iii}] doublet, solid pink line shows the determined He~{\sc ii} 
line. In top left panel, solid lines in cyan, in pink and in purple show the determined [N~{\sc ii}], [O~{\sc i}] and 
[S~{\sc ii}] doublets. In each bottom panel, horizontal dashed red lines shows residuals=$\pm$1. In each top panel, vertical 
dashed red line marks the position related to the apparent shoulder in broad Balmer line. In order to show clearer determined 
Gaussian components, the top panels are shown with y-axis in logarithmic coordinate.
}
\label{ha}
\end{figure*}

\begin{table}
\caption{Line parameters}
\begin{tabular}{llll}
\hline\hline
line & $\lambda_0$ & $\sigma$ & flux \\
\hline\hline
\multirow{2}{*}{Broad H$\alpha$}  &  6556.51$\pm$2.19  & 54.20$\pm$1.43  & 1490.49$\pm$63.93 \\
	& 6612.26$\pm$1.79 & 14.03$\pm$2.68 & 146.99$\pm$37.27 \\
\hline
\multirow{2}{*}{Broad H$\beta$}  &  4856.68$\pm$1.62  & 34.21$\pm$1.76  & 276.31$\pm$17.27 \\
        & 4897.98$\pm$1.33 & 14.90$\pm$2.11 & 55.22$\pm$12.05 \\
\hline
He~{\sc ii} & 4686.63$\pm$0.64 & 3.20$\pm$0.64 & 14.36$\pm$2.60 \\
\hline
Narrow H$\alpha$ & 6563.32$\pm$0.29 & 6.07$\pm$0.31 &  312.45$\pm$16.95 \\
\hline
Narrow H$\beta$ & 4861.92$\pm$0.29 & 4.62$\pm$0.32 &  61.32$\pm$4.74 \\
\hline
\multirow{2}{*}{[O~{\sc iii}]$\lambda5007$\AA} & 5007.72$\pm$0.03 & 3.57$\pm$0.05 & 580.08$\pm$13.21 \\
& 5005.39$\pm$0.59 & 8.65$\pm$0.69 & 99.31$\pm$12.64 \\
\hline
[O~{\sc i}]$\lambda6300$\AA & 6300.88$\pm$0.88 & 5.20$\pm$0.84 &  34.90$\pm$5.07 \\
\hline
[O~{\sc i}]$\lambda6363$\AA & 6365.12$\pm$1.56 & 8.85$\pm$1.67  & 38.02$\pm$6.48 \\
\hline
[N~{\sc ii}]$\lambda6583$\AA & 6584.72$\pm$0.24 & 5.43$\pm$0.26 & 279.21$\pm$16.53 \\ 
\hline
[S~{\sc ii}]$\lambda6716$\AA & 6716.13$\pm$1.01 & 3.82$\pm$0.82 &  51.23$\pm$16.54 \\
\hline
[S~{\sc ii}]$\lambda6731$\AA & 6729.44$\pm$1.83 & 6.63$\pm$1.68 & 71.93$\pm$17.82 \\
\hline\hline
\end{tabular}\\
Notice: The first column shows which line is measured. The Second, third, fourth columns show the measured Gaussian 
parameters: center wavelength $\lambda_0$ in units of \AA, line width (second moment) $\sigma$ in units of \AA~ and line 
flux in units of ${\rm 10^{-17}~erg/s/cm^2}$. 
\end{table}

\subsection{Spectroscopic properties of \obj}

%3.2 - 1
	Fig.~\ref{spec} shows the SDSS spectrum with PLATE-MJD-FIBERID=2200-53875-0526 of \obj~ collected from SDSS DR16 
\citep{ap21}. The apparent red-shifted shoulders, marked in Fig.~\ref{spec}, can be found in broad Balmer emission lines, 
as shown in \citet{wg21}. And the shoulders should be more apparently determined after measurements of emission lines in 
\obj~ by the following emission line fitting procedure.

%%%3.2 - 2
	Multiple Gaussian functions can be applied to well simultaneously measure the emission lines around H$\beta$ within 
rest wavelength from 4600\AA~ to 5100\AA~ and around H$\alpha$ within rest wavelength from 6150\AA~ to 6850\AA~ in \obj, 
similar as what we have recently done in \citet{zh21a, zh21b, zh21c, zh22b}. Three broad and one narrow Gaussian functions 
are applied to describe broad and narrow components in H$\alpha$ (H$\beta$). And two narrow and two broad Gaussian functions 
are applied to describe [O~{\sc iii}]$\lambda4959,5007$\AA~ doublet, for the core components and for the components related 
to shifted wings. And one Gaussian function is applied to describe He~{\sc ii}$\lambda4687$\AA~ emission line. And six Gaussian 
functions are applied to describe the [O~{\sc i}]$\lambda6300,6363$\AA, [N~{\sc ii}]$\lambda6549,6583$\AA~ and 
[S~{\sc ii}]$\lambda6716,6731$\AA~ doublets. As the following shown best fitting results, it is not necessary to consider 
components related shifted wings in the [O~{\sc i}], [N~{\sc ii}] and [S~{\sc ii}] doublets. And, a power law function is 
applied to describe the AGN continuum emissions underneath the emission lines around H$\beta$ (H$\alpha$).

%%%3.2 - 3
	When the model functions are applied, only three restrictions are accepted. First, emission flux of each Gaussian 
emission component is not smaller than zero. Second, Gaussian components of each doublet have the same redshift. Third, 
corresponding Gaussian components in broad Balmer line have the same redshift. Then, through the Levenberg-Marquardt 
least-squares minimization technique (the known MPFIT package), best fitting results to emission lines can be determined 
and shown in Fig.~\ref{ha} with $\chi^2/dof\sim1.09$. The measured parameters of each Gaussian component are listed in 
Table~3. Here, although three Gaussian functions are applied to describe each broad Balmer line, only two Gaussian components 
have reliable measured parameters which are at least 3times larger than their corresponding uncertainties, indicating two 
broad Gaussian components are enough to describe each broad Balmer line. Moreover, it is clear that there are red shifted 
shoulders in each broad Balmer emission line, especially based on the red-shifted broad Gaussian component in each broad 
Balmer line.

%%%3.2 - 4
	The main objective to discuss emission line properties of broad Balmer lines is that the shoulders can provide 
information on peak separation of two broad components coming from two BLRs related to an expected central BBH system, 
which will provide further clues on maximum space separation of the two black holes in the expected BBH system in \obj. 
The red-shifted velocity of the shoulder in broad Balmer lines are about $V_r=2200\pm300{\rm km/s}$, based on the 
red-shift broad Gaussian component in each broad Balmer line. Meanwhile, peak separation $V_p$ in units of km/s related 
to a BBH system with space separation $S$ and with BH masses of $M_{BH1}$ and $M_{BH2}$ of the two BHs can be simply 
described as 
\begin{equation}
	V_p~\sim~\sqrt{G~\times~(M_{BH1}~+~M_{BH2})/S}~\times~\sin{i}~\times~\sin(\phi)
\end{equation} 
with $i$ as inclination angle of the orbital plane to line-of-sight and $\phi$ as orientation angle of orbital phase. 
Therefore, considering the observed $V_{r}~\sim~2200{\rm km/s}~<~V_p=V_r+V_b$ ($V_b$ as blue-shifted velocity of undetected 
blue-shifted shoulders in broad Balmer lines in \obj), the maximum space separation $S$ can be simply determined as 
\begin{equation}
	S~<~\frac{G~\times~(M_{BH1}~+~M_{BH2})}{V_{p,obs}^2}
\end{equation}
. Once there are clear information of central total BH mass which will be discussed in the next section, the maximum 
space separation can be well estimated.

\subsection{Basic structure information of the expected BBH system in \obj}

%%%3.3 - 1
	For a broad line AGN, the virialization assumption accepted to broad line emission clouds is the most convenient 
method to estimate central virial BH mass as discussed in \citet{pe04, gh05, vp06, sh11}. However, considering mixed broad 
Balmer line emissions, it is difficult to determine the two clear components from central two independent BLRs related to 
central BBH system in \obj, after checking the following point.

%%%3.3 - 2
	If the determined two broad Gaussian components in each broad Balmer line shown in Fig.~\ref{ha} were truly related 
to a central BBH system, the virial BH mass as discussed in \citet{gh05} of each BH in central region can be estimated as
\begin{equation}
\begin{split}
M_{BH1}~&\propto~(f_{\alpha,r})^{0.55}(\sigma_{\alpha,r})^{2.06} \\
M_{BH2}~&\propto~(f_{\alpha,b})^{0.55}(\sigma_{\alpha,b})^{2.06}
\end{split}
\end{equation}
with $f_{\alpha,r}$ and $\sigma_{\alpha,r}$ ($f_{\alpha,b}$ and $\sigma_{\alpha,b}$) as line flux and line width (second 
moment) of red-shifted (blue-shifted) broad component in broad H$\alpha$, after considering the more recent empirical 
R-L relation to estimate BLRs sizes through line luminosity in \citet{bd13}. Then, under assumption of a BBH system, shift 
velocity ratio $R_v$ of red-shifted broad component to blue-shifted broad component in broad H$\alpha$ can be estimated as 
\begin{equation}
\begin{split}
R_v~&\sim~(\frac{f_{\alpha,b}}{f_{\alpha,r}})^{0.55}(\frac{\sigma_{\alpha,b}}{\sigma_{\alpha,r}})^{2.06} \\
	&\sim~58_{-25}^{+56}
\end{split}
\end{equation}
accepted the measured line widths and line fluxes and uncertainties listed in Table~3. 

%%%3.3 - 3
	However, according to measured central wavelengths of the two broad components in broad H$\alpha$ listed in Table~3, 
the observed shift velocity ratio $R_{v,obs}$ is 
\begin{equation}
\begin{split}
R_{v,obs}~&\sim~\frac{(6612.26\pm1.79)~-~6564.61}{6564.61~-~(6556.51\pm2.19)}\\
	&\sim~5.9_{-1.4}^{+2.5}
\end{split}
\end{equation} 
with 6564.61\ in units of \AA~ as the theoretical value of central wavelength of broad H$\alpha$ in rest frame, which is quite 
different from the $R_v$. Therefore, although there are two broad Gaussian components determined in broad H$\alpha$, the two 
components are not appropriate to be applied to determine virial BH masses of central two black holes in the expected BBH 
system. Similar results can also be found through the two components in broad H$\beta$.

%%%3.3 - 4
	Therefore, rather than virial BH mass estimated through broad line luminosity and broad line width, BH mass estimated 
by continuum luminosity as shown in \citet{pe04} is determined as 
\begin{equation}
\begin{split}
\log(\frac{M_{BH1}}{10^8{\rm M_\odot}})~&=~-0.12\pm0.07~+~(0.79\pm0.09)~\times~\log(\frac{L_1}{\rm 10^{44}}) \\
\log(\frac{M_{BH2}}{10^8{\rm M_\odot}})~&=~-0.12\pm0.07~+~(0.79\pm0.09)~\times~\log(\frac{L_2}{\rm 10^{44}})\\
\end{split}
\end{equation}
with $L_1$ and $L_2$ in units of ${\rm erg/s}$ as continuum luminosity at 5100\AA~ coming from central two BH accreting systems, 
then upper limit of central total BH mass $M_{BH}=M_{BH1}+M_{BH2}$ can be estimated as
\begin{equation}
\frac{M_{BH}}{10^8{\rm M_\odot}}~\le~10^{-0.12\pm0.07}~\times~(\frac{L_1~+~L_2}{\rm 10^{44}})^{0.79\pm0.09}
\end{equation}.
Then, based on the fitting results shown in right panels of Fig.~\ref{ha}, total continuum luminosity at 5100\AA~ in rest 
frame is $L_{t}~\sim~(1.47\pm0.02)\times10^{44}{\rm erg/s}$. Simply accepted $L_t~=~L_1~+~L_2$, upper limit of central total 
BH mass is about $(1.03\pm0.22)\times10^8{\rm M_\odot}$.

%%%3.3 - 5
	Accepted the estimated upper limit of total BH mass, the upper limit of space separation of the expected central BBH 
system in \obj~ is about 
\begin{equation}
\begin{split}
S~&<~\frac{G~\times~M_{BH}}{V_{p,obs}^2}\\
	&\sim~107\pm60 light-days 
\end{split}
\end{equation}.
Meanwhile, considering optical periodicity $\sim$340 days, as discussed in \citet{eb12}, the space separation of the central 
BBH system can be estimated as 
\begin{equation}
\begin{split}
S_{BBH}~&\sim~0.432\frac{M_{BH}}{\rm 10^8M_\odot}(\frac{T_{q}/year}{2652\frac{M_{BH}}{\rm 10^8M_\odot}})^{2/3}\\
	&\sim~2.6\pm0.3 light-days
\end{split}
\end{equation}
with accepted total BH mass $(1.03\pm0.22)\times10^8{\rm M_\odot}$. The space separation $S_{BBH}$ determined by periodicity 
is well below the upper limit of space separation determined by peak separation $V_p$, not leading to clues against assumptions 
of the central BBH system in \obj.

%%%3.3 - 6
	Before end of the subsection, one point should be noted. As simply discussed above, through optical periodicity 
$\sim$340 days, estimated space separation of the central BBH system is about 2.6 light-days in \obj. Meanwhile, continuum 
luminosity about $10^{44}{\rm erg/s}$ in \obj~ can lead BLRs sizes to be about 36 light-days through the R-L relation in 
\citet{bd13}, quite larger than $S_{BBH}\sim2.6$light-days. Therefore, there should be few effects of dynamics of central 
BBH system on emission clouds of probably mixed BLRs in \obj, or only apparent effects on emission clouds in inner regions 
of central mixed BLRs of \obj. The results can provide further clues to support that it is not appropriate to estimate 
central BH mass by properties of broad emission lines in \obj~ as well discussed above, and also to support that the shifted 
velocity ratio discussed above should be not totally confirmed to be related to dynamics of central BBH system. Multi-epoch 
monitoring of variabilities of broad emission lines should provide further and accurate properties of dynamic structures 
of central expected BBH system in \obj~ in the near future.  

\subsection{Further discussions on the origin of optical QPOs in \obj}

%%%3.4 - 1
	Meanwhile, besides the expected central BBH system, precessions of emission regions with probable hot spots for the 
optical continuum emissions can also be applied to describe the detected optical QPOs in \obj. As discussed in \citet{eh95} 
and in \citet{st03}, the expected disk precession period can be estimated as 
\begin{equation}
T_{\rm pre}\sim1040M_{8}R_{3}^{2.5}yr
\end{equation},
with $R_{3}$ as distance of optical emission regions to central BH in units of 1000 Schwarzschild radii ($R_g$) and $M_{8}$ 
as the BH mass in units of $10^8{\rm M_\odot}$. Considering optical periodicity about $T_{\rm pre}\sim340$ days and BH mass 
about $(1.03\pm0.22)\times10^8{\rm M_\odot}$ above estimated through the continuum luminosity, the expected $R_3$ could be 
around 0.06\ in \obj. 

%%%3.4 - 2
	However, based on the discussed distance of NUV emission regions to central BHs in \citet{mc10} through the 
microlensing variability properties of eleven gravitationally lensed quasars, the NUV 2500\AA~ continuum emission regions in 
\obj~ have distance from central BH as
\begin{equation}
\log{\frac{R_{2500}}{cm}}=15.78+0.80\log(\frac{M_{BH}}{10^9M_\odot})
\end{equation}
leading size of NUV emission regions to be about $60R_g$. The estimated NUV emission regions have similar distances as the 
optical continuum emission regions in \obj~ under the disk precession assumption, strongly indicating that the disk precessions 
of emission regions are not preferred to be applied to explain the detected optical QPOs in \obj.

%%%3.4 - 3
	Moreover, long-term QPOs can be detected in blazars due to jet precessions as discussed in \citet{sc18, bg19, oa20}. 
However, \obj~ is covered in Faint Images of the Radio Sky at Twenty-cm \citep{bw95, hw15}, but no apparent radio emissions. 
Therefore, jet precessions can be well ruled out to explain the optical QPOs in \obj. 

%%%3.4 - 4
	Furthermore, it is interesting to discuss whether the known relativistic Lense-Thirring precession \citep{ref1, ref2} 
can be applied to explain the detected optical QPOs in \obj. As well discussed in \citet{ref1}, observed periodicity related 
to the Lense-Thirring precession is about 
\begin{equation}
P_{LT,obs}~\sim(1+z)\times\frac{M_{BH}R_e^3}{6.45\times10^4~\times~|a_*|}
\end{equation}
with $z$ as redshift, $M_{BH}$ as BH mass in units of $\rm M_\odot$, $R_e$ in units of $R_g$ as distance of emission regions 
in central accretion disk to central BH and $a_*$ (between $\pm$1) as dimensionless BH spin parameter. If accepted central BH mass as 
$(1.03\pm0.22)\times10^8{\rm M_\odot}$, minimum value of $R_e$ as $60R_g$ (the estimated value for NUV emission regions above) 
for optical emission regions and maximum value $|a_*|=1$, the minimum $P_{LT,obs}$ can be estimated to be about 5200days, about 
13 times larger than the detected optical periodicity about 340days in \obj. Therefore, the detected optical QPOs 
in \obj~ are not related to relativistic Lense-Thirring precessions.

%%%3.4 - 5
	Before ending of the manuscript, one point is noted. The \obj~ is collected from the candidates of off-nucleus AGN 
reported in the literature. However, it is not unclear whether are there tight connections between optical QPOs and off-nucleus 
AGN, studying on a sample of off-nucleus AGN with apparent optical QPOs could provide further clues on probable intrinsic 
connections in the near future.

\section{Summary and Conclusions}
    The final summary and main conclusions are as follows. 
\begin{itemize}
\item The 4.45 years long-term ZTF g/r/i-band light curves can be well described by a sine function with periodicities 
	about 340 days (about 0.9 years) with uncertainties about 4-5 days in \obj, which can be further confirmed by the 
	corresponding sine-like phase folded light curve with accepted periodicities, indicating apparent optical QPOs 
	in \obj.
\item Confidence level higher than $5\sigma$ can be confirmed to support the optical QPOs in \obj~ through applications 
	of the Generalized Lomb-Scargle periodogram. Moreover, bootstrap method can be applied to re-determine small 
	uncertainties about 5 days of the periodicities.
\item The reliable optical QPOs with periodicities $\sim$340 days with confidence level higher than 5$\sigma$ in \obj~ 
	can also be confirmed by properties of ACF and WWZ methods, through the ZTF g/r-band light curves.  
\item Robustness of the optical QPOs in \obj~ can be confirmed by the four different methods leading to totally similar 
	periodicities with confidence level higher than 5$\sigma$.
\item Based on intrinsic AGN variabilities traced by the CAR process, confidence level higher than $3\sigma$ can be well 
	determined to support the detect optical QPOs in \obj~ are not from intrinsic AGN activities, through a sample of 
	100000 CAR process simulated light curves. Therefore, the optical QPOs in \obj~ are more confident and robust, 
	leading to an expected central BBH system in \obj.
\item Although each broad Balmer emission line can be described by two Gaussian components, shifted velocity ratio 
	determined through virial BH mass properties are totally different from the observed shifted velocity ratio 
	through the measured central wavelengths of the two broad Gaussian components, indicating that applications of 
	line parameters of the two broad Gaussian components are not preferred to estimate virial BH masses of central BHs, 
	under the assumptions of an expected central BBH system in \obj.
\item based on measured continuum luminosity at 5100\AA~ in \obj, central total BH mass can be estimated as 
	$(1.03\pm0.22)\times10^8{\rm M_\odot}$. Then, based on the apparent shoulders in broad Balmer emission lines, 
	upper limit of space separation of the expected central BBH system can be estimated as $(107\pm60)$ light-days in 
	\obj. And based on the optical periodicities, the space separation of the central BBH system can be estimated as 
	$(2.6\pm0.3)$ light-days in \obj, well below the estimated upper limit of space separation. 
\item Based on the estimated size (distance to central BH) about $60{\rm R_G}$ of the NUV emission regions similar to 
	the disk precession expected size about $60{\rm R_G}$ of the optical emission regions, the disk precessions 
	can be not preferred to explain the detected optical QPOs in \obj. 
\item There are no apparent radio emissions in \obj, strongly supporting that the jet precessions can be totally ruled 
	out to explain the detected optical QPOs in \obj.
\end{itemize}

\section*{Acknowledgements}
%Zhang gratefully acknowledge the anonymous referee for reading our manuscript carefully and patiently.
Zhang gratefully acknowledge the anonymous referee for giving us constructive comments and suggestions to greatly 
improve our paper. Zhang gratefully acknowledges the kind grant support from NSFC-12173020 and NSFC-12373014. This paper has 
made use of the data from the SDSS projects, \url{http://www.sdss3.org/}, managed by the Astrophysical Research Consortium for 
the Participating Institutions of the SDSS-III Collaboration. This paper has made use of the data from the ZTF 
\url{https://www.ztf.caltech.edu}. The paper has made use of the public JAVELIN code 
\url{(http://www.astronomy.ohio-state.edu/~yingzu/codes.html#javelin}), and the MPFIT package 
\url{https://pages.physics.wisc.edu/~craigm/idl/cmpfit.html}, and the emcee package \url{https://emcee.readthedocs.io/en/stable/}, 
and the wwz.py code \url{http://github.com/eaydin} written by M. Emre Aydin. This research has made use of the NASA/IPAC 
Extragalactic Database (NED, \url{http://ned.ipac.caltech.edu}) which is operated by the California Institute of Technology, 
under contract with the National Aeronautics and Space Administration.

\section*{Data Availability}
The data underlying this article will be shared on reasonable request to the corresponding author
(\href{mailto:xgzhang@gxu.edu.cn}{xgzhang@gxu.edu.cn}).

\label{lastpage}

\begin{thebibliography}{   }
%\bibitem[\protect\citeauthoryear{Abramowicz et al.}{2004}]{ak04}
%Abramowicz, M. A., Kluzniak, W., McClintock, J. E., Remillard, R. A., 2004, ApJL, 609, L63
\bibitem[\protect\citeauthoryear{Ahumada et al.}{2021}]{ap21}
Ahumada, R.; Prieto, C. A.; Almeida, A.; et al., 2021, ApJS, 249, 3 
\bibitem[\protect\citeauthoryear{An et al.}{2016}]{al16}
An, T.; Lu, X.; Wang, J., 2016, A\&A, 585, 89
%\bibitem[\protect\citeauthoryear{Arzoumanian et al.}{2015}]{ar15}
%Arzoumanian, Z., Brazier, A., Burke-Spolaor, S., et al., 2015, ApJ, 813, 65
\bibitem[\protect\citeauthoryear{Baldassare et al.}{2020}]{bg20}
Baldassare, V. F.; Geha, M.; Greene, J., 2020, ApJ, 896, 10
%\bibitem[\protect\citeauthoryear{Barnes \& Hernquist}{1996}]{bh96}
%Barnes, J. E.; Hernquist, L., 1996, ApJ, 471, 115
%\bibitem[\protect\citeauthoryear{Barth \& Stern}{2018}]{bs18}
%Barth, A. J.; Stern, D., 2018, ApJ, 859, 10
%\bibitem[\protect\citeauthoryear{Barth}{2015}]{bb15}
%Barth, A. J.; Bennert, V. N.; Canalizo, G.; et al., 2015, ApJS, 217, 26
\bibitem[\protect\citeauthoryear{Becker, White \& Helfand}{1995}]{bw95}
Becker, R. H., White, R. L., Helfand, D. J. 1995, ApJ, 450, 559
\bibitem[\protect\citeauthoryear{Begelman et al.}{1980}]{bb80}
Begelman, M. C., Blandford, R. D., Rees, M. J., 1980, Natur, 287, 307
\bibitem[\protect\citeauthoryear{Bellm et al.}{2019}]{bk19}
Bellm E. C., Kulkarni, S. R.; Barlow, T., et al., 2019, PASP, 131, 068003
\bibitem[\protect\citeauthoryear{Bentz et al.}{2013}]{bd13}
Bentz M. C., Denney, K. D.; Grier, C. J.; et al., 2013, ApJ, 767, 149
\bibitem[\protect\citeauthoryear{Bhatta}{2019}]{bg19}
Bhatta, G., 2019, Universe Proceedings, 17, 15, arXiv:1909.10268
\bibitem[\protect\citeauthoryear{Boroson \& Lauer}{2009}]{bl09}
Boroson, T. A., Lauer, T. R. 2009, Nature, 458, 53
\bibitem[\protect\citeauthoryear{Bottrell et al.}{2019}]{bh19}
Bottrell, C.; Hani, M. H.; Teimoorinia, H.; et al., 2019, MNRAS, 490, 5390
\bibitem[\protect\citeauthoryear{Bramich et al.}{2008}]{bv08}
Bramich, D. M.; Vidrih, S.; Wyrzykowski, L.; et al., 2008, MNRAS, 386, 887 %%stripe82
\bibitem[\protect\citeauthoryear{Bundy et al.}{2009}]{bf09}
Bundy, K.; Fukugita, M.; Ellis, R. S.; Targett, T. A.; Belli, S.; Kodama, T., 2009, ApJ, 697, 1369
\bibitem[\protect\citeauthoryear{Burke et al.}{2021}]{bs21}
Burke, C. J.; Shen, Y.; Blaes, O., et al., 2021, Sci, 373, 789
%\bibitem[\protect\citeauthoryear{Carlberg}{1992}]{cr92}
%Carlberg, R. G., 1992, ApJL, 399, 31
\bibitem[\protect\citeauthoryear{Charisi et al.}{2016}]{cb16}
Charisi, M.; Bartos, I.; Haiman, Z.; et al., 2016, MNRAS, 463, 2145
\bibitem[\protect\citeauthoryear{Chen et al.}{2022}]{ch22}
Chen, Y. C.; Hwang, H. C.; Shen, Y.; Liu, X.; Zakamska, N. L.. Yang, Q.; Li, J. I., 2022, ApJ, 925, 162
\bibitem[\protect\citeauthoryear{Comerford et al.}{2013}]{cs13}
Comerford, J. M., Schluns, K., Greene, J. E., Cool, R. J., 2013, ApJ, 777, 64
\bibitem[\protect\citeauthoryear{Cui, Zhang \& Chen}{1998}]{ref1}
Cui, W.; Zhang, S. N.; Chen, W., 1998, ApJL, 492, L53 
\bibitem[\protect\citeauthoryear{Dekany et al.}{2020}]{ds20}
Dekany, R.; Smith, R. M.; Riddle, R., et al., 2020, PASP, 132, 038001
%\bibitem[\protect\citeauthoryear{Desvignes et al.}{2016}]{de16}
%Desvignes, G., Caballero, R. N., Lentati, L., et al., 2016, MNRAS, 458, 3341
\bibitem[\protect\citeauthoryear{De Rosa et al.}{2019}]{dv19}
De Rosa, A.; Vignali, C.; Bogdanovic, T., et al., 2019, NewAR, 86, 101525
%\bibitem[\protect\citeauthoryear{Dorn-Wallenstein et al.}{2017}]{dl17}
%Dorn-Wallenstein, T.; Levesque, E. M.; Ruan, J. J., 2017, ApJ, 850, 86 
\bibitem[\protect\citeauthoryear{Drake et al.}{2009}]{dr09}
Drake, A. J.; Djorgovski, S. G.; Mahabal, A., et al., 2009, ApJ, 696, 870
%\bibitem[\protect\citeauthoryear{Eracleous \& Halpern}{1994}]{eh94}
%Eracleous, M.; Halpern, J. P., 1994, ApJS, 90, 1
\bibitem[\protect\citeauthoryear{Eracleous et al.}{1995}]{eh95}
Eracleous, M.; Livio M.; Halpern, J. P., 1995, ApJ, 438, 610
\bibitem[\protect\citeauthoryear{Eracleous et al.}{2012}]{eb12}
Eracleous, M.; Boroson, T. A.; Halpern, J. P.; Liu, J., 2012, ApJS, 201, 23
\bibitem[\protect\citeauthoryear{Flewelling et al.}{2020}]{fm20}
Flewelling, H. A.; Magnier, E. A.; Chambers, K. C.; et al., 2020, ApJS, 251, 7
\bibitem[\protect\citeauthoryear{Foreman-Mackey et al.}{2013}]{fh13}
Foreman-Mackey, D.; Hogg, D. W.; Lang, D.; Goodman, J., 2016, PASP, 125, 306
%\bibitem[\protect\citeauthoryear{Foster \& Backer}{1990}]{fb90}
%Foster, R. S., \& Backer, D. C. 1990, ApJ, 361, 300
\bibitem[\protect\citeauthoryear{Foster}{1996}]{fg96}
Foster, G., 1996, AJ, 112, 1709
\bibitem[\protect\citeauthoryear{Fragione et al.}{2019}]{fg19}
Fragione, G.; Grishin, E.; Leigh, N. W. C.; Perets, H. B.; Perna, R., 2019, MNRAS, 488, 47
%\bibitem[\protect\citeauthoryear{Gaskell}{2010}]{gm10}
%Gaskell, C. M., 2010, Natur, 463, 1
%\bibitem[\protect\citeauthoryear{Gierlinski et al.}{2008}]{gm08}
%Gierlinski, M.; Middleton, M.; Ward, M.; Done, C., 2008, Nature, 455, 369
\bibitem[\protect\citeauthoryear{Graham et al.}{2015a}]{gd15a}
Graham, M. J.; Djorgovski, S. G.; Stern, D., et al., 2015a, Natur, 518, 74
\bibitem[\protect\citeauthoryear{Graham et al.}{2015}]{gd15}
Graham, M. J., Djorgovski, S. G., Stern, D., et al., 2015, MNRAS, 453, 1562
%\bibitem[\protect\citeauthoryear{Greene \& Ho}{2005a}]{gh05a}
%Greene, J. E.; Ho, L. C., 2005a, ApJ, 627, 721
\bibitem[\protect\citeauthoryear{Greene \& Ho}{2005}]{gh05}
Green, J. E., Ho, L. C., 2005, ApJ, 630, 122
\bibitem[\protect\citeauthoryear{Gupta et al.}{2018}]{gt18}
Gupta, A. C.; Tripathi, A.; Wiita, P. J.; et al., 2018, A\&A, 616, 6
\bibitem[\protect\citeauthoryear{Helfand et al.}{2015}]{hw15}
Helfand, D. J.; White, R. L.; Becker, R. H., 2015, ApJ, 801, 26
\bibitem[\protect\citeauthoryear{Hinshaw et al.}{2013}]{hl13}
Hinshaw, G.; Larson, D.; Komatsu, E.; et al., 2013, ApJS, 208, 19
%\bibitem[\protect\citeauthoryear{Kim et al.}{2018}]{km18}
%Kim, D. C.; Yoon, I.; Evans, A. S., 2018, ApJ, 861, 51
%\bibitem[\protect\citeauthoryear{Ingram \& Motta}{2020}]{im20}
%Ingram, A.; Motta, S., 2020, New Astronomy Reviews, arXiv:2001.08758
\bibitem[\protect\citeauthoryear{Ivezic et al.}{2019}]{ic19}
Ivezic, Z.; Connolly, A. J.; VanderPlas, J. T.; Gray, A. 2019, Statistics, Data Mining, 
and Machine Learning in Astronomy: A Practical Python Guide for the Analysis of Survey Data, ISBN: 9780691197050, 
Princeton University Press
%\bibitem[\protect\citeauthoryear{James et al.}{2010}]{jp10}
%James, M; Paul, B.; Devasia, J.; Indulekha, K., 2010, MNRAS, 407, 285
%\bibitem[\protect\citeauthoryear{Jin et al.}{2021}]{jd21}
%Jin, C.; Done, C.; Ward, M., 2021, MNRAS, 500, 2475
\bibitem[\protect\citeauthoryear{Kauffmann et al.}{1993}]{kw93}
Kauffmann, G.; White, S. D. M.; Guiderdoni, B., 1993, MNRAS, 264, 201
\bibitem[\protect\citeauthoryear{Kelly, Bechtold \& Siemiginowska}{2009}]{kbs09}
Kelly, B. C.; Bechtold, J.; Siemiginowska, A., 2009, ApJ, 698, 895
%\bibitem[\protect\citeauthoryear{Kelly et al.}{2014}]{kb14}
%Kelly, B. C.; Becker, A. C.; Sobolewska, M.; Siemiginowska, A.; Uttley, P., 2014, ApJ, 788, 33
\bibitem[\protect\citeauthoryear{Kochanek et al.}{2017}]{ks17}
Kochanek, C. S.; Shappee, B. J.; Stanek, K. Z.; et al., 2017, PASP, 129, 4502
\bibitem[\protect\citeauthoryear{Kollatschny et al.}{2020}]{kw20}
Kollatschny, W.; Weilbacher, P. M.; Ochmann, M. W.; Chelouche, D.; Monreal-Ibero, A.; Bacon, R.; 
Contini, T., 2020, A\&A, 633, 79
\bibitem[\protect\citeauthoryear{Komossa et al.}{2003}]{km03}
Komossa, S., Burwitz, V., Hasinger, G., Predehl, P., Kaastra, J. S., Ikebe, Y., 2003, ApJL, 582, L15
\bibitem[\protect\citeauthoryear{Komossa et al.}{2008}]{kz08}
Komossa, S., Zhou, H., Lu, H. 2008, ApJ, 678, L81
%\bibitem[\protect\citeauthoryear{Kormendy et al.}{2009}]{kf09}
%Kormendy, J.; Fisher, D. B.; Cornell, M. E.; Bender, R., 2009, ApJS, 182, 216
\bibitem[\protect\citeauthoryear{Kovacevic et al.}{2019}]{kp19}
Kovacevic, A. B., Popovic, L. C., Simic, S., Ilic, D., 2019, ApJ, 871, 32
\bibitem[\protect\citeauthoryear{Kovacevic et al.}{2020}]{ky20}
Kovacevic, A. B.; Yi, T.; Dai, X.; et al., 2020, MNRAS, 494, 4069
\bibitem[\protect\citeauthoryear{Kozlowski et al.}{2010}]{koz10}
Kozlowski, S.; Kochanek, C. S.; Udalski, A., et al., 2010, ApJ, 708, 927
\bibitem[\protect\citeauthoryear{Krishnan et al.}{2021}]{km21}
Krishnan, S.; Markowitz, A. G.; Schwarzenberg-Czerny, A.; Middleton, M. J., 2021, MNRAS, 508, 3975
\bibitem[\protect\citeauthoryear{Kushwaha et al.}{2020}]{ks20}
Kushwaha, P.; Sarkar, A.; Gupta, Alok C.; Tripathi, A.; Wiita, P. J., 2020, MNRAS, 499, 653
%\bibitem[\protect\citeauthoryear{Lacey \& Cole}{1993}]{lc93}
%Lacey, C.; Cole, S., 1993, MNRAS, 262, 627
\bibitem[\protect\citeauthoryear{Li et al.}{2021}]{ly21}
Li, X.; Cai, Y.; Yang, H.; Luo, Y.; Yan, Y.; He, J.; Wang, L., 2021, MNRAS, 506, 2540
\bibitem[\protect\citeauthoryear{Liao et al.}{2021}]{lw21}
Liao, W.; Chen, Y.; Liu, X.; et al., 2021, MNRAS, 500, 4025
%\bibitem[\protect\citeauthoryear{Lin et al.}{2013}]{li13}
%Lin, D. C.; Irwin, J. A.; Godet, O.; Webb, N. A.; Barret, D., 2013, ApJL, 776
\bibitem[\protect\citeauthoryear{Lin et al.}{2004}]{lk04}
Lin, L.; Koo, D. C.; Willmer, C. N. A.; et al., 2004, ApJL, 617, 9
\bibitem[\protect\citeauthoryear{Liu et al.}{2016}]{le16}
Liu, J., Eracleous, M., Halpern, J. P., 2016, ApJ, 817, 42
\bibitem[\protect\citeauthoryear{Liu et al.}{2015}]{lg15}
Liu, T., Gezari, S., Heinis, S., et al., 2015, ApJL, 803, L16
\bibitem[\protect\citeauthoryear{Liu et al.}{2018}]{lg18}
Liu, T., Gezari, S., Miller M. C., 2018, ApJL, 859, L12
\bibitem[\protect\citeauthoryear{Lomb}{1976}]{ln76}
Lomb, N. R. 1976, Ap\&SS, 39, 447
\bibitem[\protect\citeauthoryear{MacLeod et al.}{2010}]{mi10}
MacLeod, C. L., Ivezic, Z., Kochanek, C. S., et al., 2010, ApJ, 721, 1014
\bibitem[\protect\citeauthoryear{Madejski \& Sikora}{2016}]{ms16}
Madejski, G.; Sikora, M., 2016, ARA\&A, 54, 725
\bibitem[\protect\citeauthoryear{Magnier et al.}{2020}]{mc20}
Magnier, E. A.; Chambers, K. C.; Flewelling, H. A.; et al., 2020, ApJS, 251, 3
\bibitem[\protect\citeauthoryear{Mannerkoski et al.}{2022}]{mj22}
Mannerkoski, M.; Johansson, P. H.; Rantala, A.; Naab, T.; Liao, S.; Rawlings, A., 2022, ApJ, 929, 167
%\bibitem[\protect\citeauthoryear{Markowitz, Reeves \& Braito}{2016}]{mr16}
%Markowitz, A.; Reeves, J. N.; Braito, V., 2006, ApJ, 646,783
\bibitem[\protect\citeauthoryear{Markwardt}{2009}]{mpf09} 
Markwardt, C. B., 2009, Astronomical Data Analysis Software and Systems XVIII ASP Conference Series, Vol. 411, 
proceedings of the conference held 2-5 November 2008 at Hotel Loews Le Concorde, Quebec City, QC, Canada. Edited 
by David A. Bohlender, Daniel Durand, and Patrick Dowler. San Francisco: Astronomical Society of the Pacific, p.251
\bibitem[\protect\citeauthoryear{Martin et al.}{2021}]{mj21}
Martin, G.; Jackson, R. A.; Kaviraj, S., et al., 2021, MNRAS, 500, 4937
\bibitem[\protect\citeauthoryear{Mayer et al.}{2010}]{mk10}
Mayer, L., Kazantzidis, S., Escala, A., Callegari, S., 2010, Natur, 466, 1082
%\bibitem[\protect\citeauthoryear{Mayer}{2013}]{ml13}
%Mayer, L., 2013, Classical and Quantum Gravity, 30 244008
%\bibitem[\protect\citeauthoryear{Mayer \& Bonoli}{2019}]{mb19}
%Mayer, L., Bonoli, S., 2019, Reports on Progress in Physics, 82, 016901
%\bibitem[\protect\citeauthoryear{McHardy et al.}{2006}]{mk06}
%McHardy, I. M.; Koerding, E.; Knigge, C.; Uttley, P.; Fender, R. P., 2006, Nature, 444, 7
%\bibitem[\protect\citeauthoryear{Menou et al.}{2001}]{mh01}
%Menou, K.; Haiman, Z.; Narayanan, V. K., 2001, ApJ, 558, 535
\bibitem[\protect\citeauthoryear{Merritt}{2006}]{md06}
Merritt, D., 2006, ApJ, 648, 976
\bibitem[\protect\citeauthoryear{Morgan et al.}{2010}]{mc10}
Morgan, C. W.; Kochanek, C. S.; Morgan, N. D.; Falco, E. E., 2010, ApJ, 712, 1129
\bibitem[\protect\citeauthoryear{Moreno et al.}{2019}]{mv19}
Moreno, J.; Vogeley, M. S.; Richards, G. T.; Yu, W., 2019, PASP, 131, 3001
\bibitem[\protect\citeauthoryear{Nardini}{2017}]{ne17}
Nardini, E., 2017, MNRAS, 471, 3483
%\bibitem[\protect\citeauthoryear{Obric et al.}{2006}]{oi06}
%Obric, M.; Ivezic, Z.; Best, P. N., et al., 2006, MNRAS, 370, 1677
\bibitem[\protect\citeauthoryear{Otero-Santos et al.}{2020}]{oa20}
Otero-Santos, J.; Acosta-Pulido, J. A.; Becerra Gonzalez, J.; et al., 2020, MNRAS, 492, 5524
%\bibitem[\protect\citeauthoryear{Papadakis \& Lawrence}{1993}]{pl93}
%Papadakis, I. E.; Lawrence, A., 1993, Natur, 361, 233
%\bibitem[\protect\citeauthoryear{Paris et al.}{2018}]{pa18}
%Paris, I.; Petitjean, P.; Aubourg, E.; et al., 2018, A\&A, 613, 51
%\bibitem[\protect\citeauthoryear{Pasham, Strohmayer \& Mushotzky}{2014}]{ps14}
%Pasham, D. R.; Strohmayer, T. E.; Mushotzky, R. F.; 2014, Nature, 513, 74
\bibitem[\protect\citeauthoryear{Penil1 et al.}{2020}]{pd20}
Penil1, P.; Dominguez1, A.; Buson, S.; et al., 2020, ApJ, 896, 134
\bibitem[\protect\citeauthoryear{Peterson et al.}{2004}]{pe04}
Peterson B. M.; Ferrarese, L.; Gilbert, K. M., et al., 2004, ApJ, 613, 682
\bibitem[\protect\citeauthoryear{Piconcelli et al.}{2010}]{pv10}
Piconcelli, E., Vignali, C., Bianchi, S., et al., 2010, ApJL 722, L147
%\bibitem[\protect\citeauthoryear{Reardon et al.}{2016}]{re16}
%Reardon, D. J., Hobbs, G., Coles, W., et al., 2016, MNRAS, 455, 1751
%\bibitem[\protect\citeauthoryear{Reines et al.}{2016}]{rr16}
%Reines, A. E.; Reynolds, M. T.; Miller, J. M.; et al., 2016, ApJL, 830, L35
%\bibitem[\protect\citeauthoryear{Remillard \& McClintock}{2006}]{rm06}
%Remillard, R. A.; McClintock, J. E., 2006, ARA\&A, 44, 49
\bibitem[\protect\citeauthoryear{Rees}{1984}]{mr84}
Rees, M. J., 1984, ARA\&A, 22, 471
\bibitem[\protect\citeauthoryear{Rodriguez et al.}{2009}]{rt09}
Rodriguez, C., Taylor, G. B., Zavala, R. T., Pihlstrom, Y. M., Peck, A. B., 2009, ApJ, 697, 37
%\bibitem[\protect\citeauthoryear{Rodriguez-Gomez et al.}{2016}]{rp16}
%Rodriguez-Gomez, V.; Pillepich, A.; Sales, L. V., et al., 2016, MNRAS, 458, 2371
\bibitem[\protect\citeauthoryear{Rodriguez-Gomez et al.}{2017}]{rs17}
Rodriguez-Gomez, V.; Sales, L. V.; Genel, S., et al., 2017, MNRAS, 467, 3083
\bibitem[\protect\citeauthoryear{Sandrinelli et al.}{2018}]{sc18}
Sandrinelli, A.; Covino, S.; Treves, A.; et al., 2018, A\&A, 615, 118
\bibitem[\protect\citeauthoryear{Satyapal et al.}{2014}]{se14}
Satyapal, S.; Ellison, S. L.; McAlpine, W.; Hickox, R. C.; Patton, D. R.; Mendel, J. T., 2014, MNRAS, 441, 1297
\bibitem[\protect\citeauthoryear{Scargle}{1982}]{sj82}
Scargle, J. D. 1982, ApJ, 263, 835
\bibitem[\protect\citeauthoryear{Sesana et al.}{2018}]{se18}
Sesana, A.; Haiman, Z.; Kocsis, B.; Kelley, L. Z., 2018, ApJ, 856, 42
\bibitem[\protect\citeauthoryear{Serafinelli et al.}{2020}]{ss20}
Serafinelli, R.; Severgnini, P.; Braito, V., et al., 2020, ApJ, 902, 10
\bibitem[\protect\citeauthoryear{Shappee et al.}{2014}]{sp14}
Shappee, B. J.; Prieto, J. L.; Grupe, D.; et al., 2014, ApJ, 788, 48
\bibitem[\protect\citeauthoryear{Shen \& Loeb}{2010}]{sl10}
Shen, Y.; Loeb, A., 2010, ApJ, 725, 249
\bibitem[\protect\citeauthoryear{Shen et al.}{2011}]{sh11}
Shen, Y.; Richards, G. T.; Strauss, M. A; et al., 2011, ApJS, 194, 4
%\bibitem[\protect\citeauthoryear{Sazonov et al.}{2021}]{sg21}
%Sazonov, S.; Gilfanov, M.; Medvedev, P.; et al., 2021, MNRAS, 508, 3820
\bibitem[\protect\citeauthoryear{Sheng, Ross \& Nicholl}{2022}]{sr22}
Sheng, X.; Ross, N.; Nicholl, M., 2022, MNRAS, 512, 5580
\bibitem[\protect\citeauthoryear{Silk \& Rees}{1998}]{sr98}
Silk, J.; Rees, M. J. 1998, A\&A, 331, L1
\bibitem[\protect\citeauthoryear{Smith et al.}{2009}]{ss09}
Smith, K. L., Shields, G. A., Bonning, E. W., McMullen, C. C., Salviander, S. 2009, ApJ, 716, 866
%\bibitem[\protect\citeauthoryear{Smith et al.}{2018}]{sm18}
%Smith, K. L.; Mushotzky, R. F.; Boyd, P. T.; Wagoner, R. V., 2018, ApJL, 860, L10
%\bibitem[\protect\citeauthoryear{Songshen et al.}{2020}]{sx20}
%Songsheng, Y.; Xiao, M.; Wang, J.; Ho, L. C., 2020, ApJS, 247, 3
%\bibitem[\protect\citeauthoryear{Starkey et al.}{2016}]{sh16}
%Starkey, D. A.; Horne, K.; Villforth, C., 2016, MNRAS, 456, 1960
\bibitem[\protect\citeauthoryear{Storchi-Bergmann et al.}{2003}]{st03}
Storchi-Bergmann, T.; Nemmen da Silva, R., Eracleous, M., et al., 2003, ApJ, 598, 956
\bibitem[\protect\citeauthoryear{Takata, Mukuta \& Mizumoto}{2018}]{tm18}
Takata, T.; Mukuta, Y.; Mizumoto, Y., 2018, ApJ, 869, 178
\bibitem[\protect\citeauthoryear{Thanjavur et al.}{2021}]{ti21}
Thanjavur, K.; Ivezic, Z.; Allam, S. S.; Tucker, D. L.; Smith, J. A.; Gwyn, S., 2021, MNRAS, 505, 5941
\bibitem[\protect\citeauthoryear{Ulrich et al.}{1997}]{um97}
Ulrich, M. H.; Maraschi, L.; Urry, C. M., 1997, ARA\&A, 35, 445
%\bibitem[\protect\citeauthoryear{van den Eijnden et al.}{2017}]{vi17}
%van den Eijnden, J.; Ingram, A.; Uttley, P.; et al., 2017, MNRAS, 464, 2643
%\bibitem[\protect\citeauthoryear{van der Klis}{2000}]{vm00}
%van der Klis, M., 2000, ARA\&A, 38, 717
%\bibitem[\protect\citeauthoryear{van der Klis}{1989}]{vm98}
%van der Klis, M., 1989, ARA\&A, 27, 517
\bibitem[\protect\citeauthoryear{VanderPlas}{2018}]{vj18}
VanderPlas, J. T., 2018, ApJS, 236, 16 
\bibitem[\protect\citeauthoryear{Vaughan et al.}{2016}]{vu16}
Vaughan, S.; Uttley, P.; Markowitz, A. G.; et al., 2016, MNRAS, 461, 3145
%\bibitem[\protect\citeauthoryear{Verbiest et al.}{2016}]{ve16}
%Verbiest, J. P. W., Lentati, L., Hobbs, B., et al., 2016, MNRAS, 458, 1267
\bibitem[\protect\citeauthoryear{Vestergaard \& Peterson}{2006}]{vp06}
Vestergaard, M., Peterson, B. M. 2006, ApJ, 641, 689
\bibitem[\protect\citeauthoryear{Ward et al.}{2021}]{wg21}
Ward, C.; Gezari, S.; Frederick, S., et al., 2021, ApJ, 913, 102
%\bibitem[\protect\citeauthoryear{Wagoner}{2012}]{ref2}
%Wagoner, R. V., 2012, ApJL, 752, L18
\bibitem[\protect\citeauthoryear{Wang et al.}{2017}]{wg17}
Wang, L., Greene, J. E., Ju, W., Rafikov, R. R., Ruan J. J., Schneider, D. P., 2017, ApJ, 834, 129
\bibitem[\protect\citeauthoryear{Wang et al.}{2023}]{ws23}
Wang, J.; Songsheng, Y.; Li, Y.; Du, P., 2023, MNRAS, 518, 3397
\bibitem[\protect\citeauthoryear{Wagoner}{2012}]{ref2}
	Wagoner, R. V., 2012, ApJL, 752, L18
\bibitem[\protect\citeauthoryear{Yoon et al.}{2022}]{yp22}
Yoon, Y.; Park, C.; Chung, H.; Lane, R. R., 2022, ApJ, 925, 168 
\bibitem[\protect\citeauthoryear{Zechmeister \& Kurster}{2009}]{zk09}
Zechmeister, M.; Kurster, M., 2009, A\&A, 496, 577
\bibitem[\protect\citeauthoryear{Zhang \& Feng}{2017}]{zh17}
Zhang, X. G., Feng L. L., 2017, MNRAS, 464, 2203
\bibitem[\protect\citeauthoryear{Zhang}{2021a}]{zh21a}
Zhang, X. G., 2021a, MNRAS, 502, 2508 
\bibitem[\protect\citeauthoryear{Zhang}{2021b}]{zh21b}
Zhang, X. G., 2021b, ApJ, 909, 16, arXiv2101.02465
\bibitem[\protect\citeauthoryear{Zhang}{2021c}]{zh21c}
Zhang, X. G., 2021c, ApJ accepted, arXiv2107.09214
\bibitem[\protect\citeauthoryear{Zhang}{2021d}]{zh21d}
Zhang, X. G., 2021d, MNRAS, 507, 5205,  arXiv:2108.09714
\bibitem[\protect\citeauthoryear{Zhang}{2022a}]{zh22a}
Zhang, X. G., 2022a, MNRAS, 512, 1003, arXiv:2202.11995
\bibitem[\protect\citeauthoryear{Zhang \& Zhao}{2022b}]{zh22b}
Zhang, X. G., Zhao S., 2022b, ApJ, 937, 105, ArXiv:2209.02164  %%%True Type-2 QSO
\bibitem[\protect\citeauthoryear{Zhang}{2022c}]{zh22c}
Zhang, X. G., 2022c, MNRAS, 516, 3650, arXiv:2209.01923  %%%%QPO combine CSS and ZTF
\bibitem[\protect\citeauthoryear{Zheng et al.}{2016}]{zb16}
Zheng, Z.; Butler, N. R.; Shen, Y.; et al., 2016, ApJ, 827, 56
\bibitem[\protect\citeauthoryear{Zhou et al.}{2004}]{zw04}
Zhou, H., Wang, T., Zhang, X., Dong, X., Li, C. 2004, ApJL, 604, L33
\bibitem[\protect\citeauthoryear{Zu et al.}{2013}]{zk13}
Zu, Y.; Kochanek, C. S.; Kozlowski, S.; Udalski, A., 2013, ApJ, 765, 106
\bibitem[\protect\citeauthoryear{Zu et al.}{2016}]{zk16}
Zu, Y.; Kochanek, C. S.; Kozlowski, S.; Peterson, B. M., 2016, ApJ, 819, 122
\end{thebibliography}
\end{document}